\documentclass[%
 reprint,
 amsmath,amssymb,
 aps,
]{revtex4-2}

\usepackage{todonotes}
\usepackage{graphicx}% Include figure files
\usepackage{dcolumn}% Align table columns on decimal point
\usepackage{bm}
\usepackage{physics}
\usepackage{amsmath}
\usepackage{amssymb}
\usepackage{bbold}
\usepackage{tablefootnote}
\usepackage{todonotes}
\usepackage{xcolor}% bold math
\usepackage{hyperref}% add hypertext capabilities
\begin{document}

\preprint{APS/123-QED}
\title{Quantum Simulation of an Open System: \\
 A Dissipative 1+1D Ising Model}

\author{\vspace{0.6cm}Erik Gustafson$^{a}$, Michael Hite$^b$, Jay Hubisz$^c$, Bharath Sambasivam$^c$, Judah Unmuth-Yockey$^a$}

\affiliation{\vspace{0.6cm}$^a$Department of Theoretical Physics, Fermi National Accelerator Laboratory, Batavia, IL 60510}\vspace{0.6cm}
\affiliation{\vspace{0.2cm}$^b$University of Iowa Department of Physics and Astronomy, Iowa City, IA 52242}%
\affiliation{\vspace{0.2cm}$^c$Department of Physics, Syracuse University, Syracuse, NY 13244\vspace{0.6cm}}

\date{\today}% It is always \today, today,
             %  but any date may be explicitly specified

\begin{abstract}

The 1+1D Ising model is an ideal benchmark for quantum algorithms, as it is very well understood theoretically.  This is true even when expanding the model to include complex coupling constants.  In this work, we implement quantum algorithms designed for the simulation of open or complex coupling quantum field theories on IBM devices with a focus on the measurement of the Lee-Yang edge singularity.  This feature corresponds (at large volumes) to a phase transition, and our successful reproduction of the transition represents a non-trivial test for current hardware and its ability to distinguish features of interest in quantum field theories.
\end{abstract}

%\keywords{Suggested keywords}%Use showkeys class option if keyword
                              %display desired
\maketitle

%%%%%%%%%%%%%%%%%%%%%%%%%%%%%%%%%%%%%%%%%%%%%%%%%%%%%%%%%%%%%%%%%%%%%%%%%%%%%%%%
%%%%%%%%%%%%%%%%%%%%%%%%%%%%%%%%%%%%%%%%%%%%%%%%%%%%%%%%%%%%%%%%%%%%%%%%%%%%%%%%

\section{\label{sec:introduction} Introduction}

With the development of quantum algorithms designed for the simulation of lattice quantum field theory, it is of crucial importance to perform benchmarking tests. These tests should push the limits of current quantum hardware, and ideally inspire confidence that features in simulation output correspond to physics of interest, as opposed to noise in the hardware.  Quantum simulating simple, well-understood theories on contemporary noisy devices is thus useful for two main reasons: 
\begin{itemize}
    \item[1.] It reveals the amount of resources required to simulate them and motivates the construction of more resource-efficient digitization methods.
    \item[2.] It teaches us about the types of noise in NISQ-era devices, and enables us to compare various error mitigation methods.
\end{itemize}
The 1+1D Ising model is very well understood theoretically, even in the complex plane of coupling parameters, and has a rich phenomenology.  It is thus an excellent model for benchmarking, and is the model of focus in our work.

Dissipative models are particularly interesting due to their role in characterizing the behavior of effective theories where some fraction of a complete system has been traced out, such as in cosmological scenarios, or in the case of a thermal bath or chemical potential.  They are typically difficult to study using classical monte-carlo methods due to the sign problem.  In this work, we consider the well-studied (see, e.g. \cite{PhysRevLett.43.805,itzykson_drouffe_1989,UZELAC19801011,GvonGehlen_1991}) 1+1D transverse quantum Ising model with imaginary longitudinal field, where a dissipative term in an effective Hamiltonian has been included:
\begin{equation}\label{eq:Ham}
    \hat{H} = -\lambda \sum_{i=1}^{N_s - 1} \hat{\sigma}^z_i \hat{\sigma}^z_{i + 1} - h_x \sum_{i = 1}^{N_s} \hat{\sigma}^x_i + i \Theta \sum_{i = 1}^{N_s} \hat{\sigma}^z_i.
\end{equation}
In quantum mechanics, this sort of effective Hamiltonian can be understood as characterizing the time evolution of an incomplete Hilbert space, and the time evolution of a system according to this Hamiltonian must be completed into a trace-preserving set of quantum operations.

In this paper, we focus our study on the quantum simulation of exceptional points of this effective Hamiltonian, which are associated with the Lee-Yang edge singularity in the infinite volume limit \cite{GvonGehlen_1991}.  Exceptional points correspond to a reduction in the effective rank of the system at which the Hamiltonian is diagonalizable only up to Jordan-block form. The study of non-Hermitian Hamiltonians was first made popular by Bender et. al. via the study of PT-symmetric Hamiltonians \cite{PhysRevLett.80.5243,10.1063/1.532860}. More recently, they have been demonstrated experimentally \cite{Guo:2009yqd}.

Over the past decade, quantum computers have become more accessible through cloud-based platforms like IBM Quantum. There has been significant progress in the reduction of qubit and gate noise, and the improvement of coherence times; however, the level of noise is not low enough to be able to implement quantum error correction successfully. That being said, there have been great strides in understanding the types of errors in hardware, converting some types to another, and development and implementation of various error mitigation strategies. Some well-studied examples are dynamical decoupling \cite{PhysRev.80.580, ezzell2023dynamical}, measurement error mitigation \cite{Bravyi_2021, PRXQuantum.2.040326,nachman2020unfolding}, Pauli twirling \cite{PhysRevA.54.3824,PhysRevLett.76.722}, zero-noise extrapolation \cite{PhysRevLett.119.180509,PhysRevX.7.021050,Kandala2019}, mitigation circuits \cite{Urbanek_2021}, and techniques involving machine learning \cite{PhysRevLett.119.180509,OBrien2023,Strikis_2021}. See \cite{Qin2023} for a statistical analysis and scaling of error mitigation techniques with system size. Some of these protocols are now integrated into \texttt{qiskit}~\cite{Qiskit} and other platforms, making it easy to choose the level of error mitigation while implementing models. This has enabled the implementation of the dynamics and ground state preparation of physical theories using longer circuit depths than previously possible. 

In this work, we show that, using current error mitigation techniques, the exceptional points of the Hamiltonian in Eq.~\ref{eq:Ham} can be observed in quantum simulation on IBM hardware for the 1D Ising chain with an imaginary longitudinal magnetic field for small system sizes.

There exist several algorithms for simulating non-Hermitian Hamiltonians \cite{Su_2020,guimaraes2023noiseassisted,Motta2019,xie2023variational,Zhang_2022,Liu_2023,LCU,schlimgen2022quantum,Wei2016,Hu2020,PhysRevA.104.052420,Hu2022generalquantum,PhysRevResearch.3.013182}. For example, quantum imaginary time evolution (QITE) \cite{Motta2019} is a hybrid quantum NISQ-friendly algorithm that simulates imaginary time evolution of a Hamiltonian using a classical optimization technique to find a unitary that performs the correct Trotter evolution of the system. This is followed by state tomography, and then evolution again. Close to regions of criticality, when the correlation length diverges, tomography is very expensive, so QITE is not very effective near such interesting critical features. There has also been some work done on the variational quantum eigensolver~\cite{xie2023variational, Zhang_2022, Liu_2023} techniques for non-Hermitian Hamiltonians. The linear combination of unitaries~\cite{LCU} algorithm can simulate non-unitary dynamics up to controllable tolerance, but requires many ancillary qubits and oracles, and is not suited for NISQ-era implementation.

There are a set of probabilistic algorithms that make use of the quantum channels formalism~\cite{Wei2016,Hu2020,PhysRevA.104.052420,Hu2022generalquantum,PhysRevResearch.3.013182}. In this paper, we implement one of the algorithms in \cite{PhysRevA.104.052420} (the damping channel) on real quantum hardware at IBM, and demonstrate that the interesting features of non-unitary dynamics can be observed even on noisy hardware. The broad idea behind this quantum channel is to expand the system of interest using ancillary qubits. The extended system has unitary dynamics, and careful measurements on the ancillary qubits evolves the system of interest under the non-Hermitian Hamiltonian of interest with a certain probability of success. Non-Hermitian Hamiltonians have been studied using other methods on quantum hardware in Refs.~\cite{Naghiloo2019-wh,Dogra2021-ln,Lin2022,Powers2023-en}.

%{JH:  We should move this paragraph to results section, or if it is said there, just remove it here...}
%Given topological constraints of the IBM quantum hardware, we stick to the case of the imaginary magnetic field acting on just the alternate sites. We do this to avoid needing to apply additional (expensive) SWAP gates to couple each site to an ancillary qubit.  This modified model still has exceptional points as we will see in Sec.~\ref{sec:Model}, so it will be a good benchmark to test the algorithm. We simulate this model on two, four, and six sites and measure a diagonal observable that is sensitive to the exceptional line.

The paper is organized as follows: In section \ref{sec:Theory}, we review the damping channel used to simulate non-Hermitian dynamics using ancillary qubits. In \ref{sec:Model}, we provide the Hamiltonian for the model we study in this work, and review the unique features of non-Hermitian Hamiltonians we are looking to observe, i.e., the exceptional points, and showcase an example of the 1D Ising chain with an imaginary magnetic field. In \ref{sec:Results}, we briefly review the error mitigation strategies we use, and present the results from quantum simulation of the aforementioned modified Ising model. Finally, in \ref{sec:Conclusions}, we conclude and offer further directions to pursue.

%%%%%%%%%%%%%%%%%%%%%%%%%%%%%%%%%%%%%%%%%%%%%%%%%%%%%%%%%%%%%%%%%%%%%%%%%%%%%%%%
%%%%%%%%%%%%%%%%%%%%%%%%%%%%%%%%%%%%%%%%%%%%%%%%%%%%%%%%%%%%%%%%%%%%%%%%%%%%%%%%

\section{The Damping Channel}
\label{sec:Theory}

In this section, we review the quantum channel used to perform a Trotter step according to the anti-Hermitian part of the Hamiltonian.
The  full details of the channel can be found in {\color{blue} Ref} \cite{PhysRevA.104.052420}. This channel simulates a single qubit anti-Hermitian Hamiltonian 
\begin{equation}\label{eq:SingNHHam}
    \hat{k}_j=i\Theta\,\hat{\sigma}_z,    
\end{equation}
where $j$ is the index of the qubit this term is acting on. Any single qubit anti-Hermitian Hamiltonian can be brought to this form using simple rotations. Moreover, if the anti-Hermitian Hamiltonian of interest acts on multiple qubits, under the Trotter approximation, it can be decomposed into a unitary acting on the relevant qubits and the non-unitary piece acting on just a single qubit~\cite{PhysRevA.104.052420}. The Kraus operator set that defines the damping channel is
\begin{equation}
    \hat{E}_0^{\text{DC}}=\begin{pmatrix}
        1 & 0 \\
        0 & \sqrt{1-\gamma}
    \end{pmatrix}
    \hspace{0.5cm}\text{and}\hspace{0.5cm}
    \hat{E}_1^{\text{DC}}=\begin{pmatrix}
        0 & 0 \\
        0 & \sqrt{\gamma}
    \end{pmatrix},
\end{equation}
where $\gamma = 1-e^{-4\Theta\delta t}$. These Kraus operators satisfy the usual completeness relation
\begin{equation}
    \hat{E}_0^{\text{DC}\dagger}\hat{E}_0^{\text{DC}}+\hat{E}_1^{\text{DC}\dagger}\hat{E}_1^{\text{DC}}=\mathbb{1}.
\end{equation}
The first Kraus operator, $\hat{E}_0^{\text{DC}}$ is defined such that its action on the system qubit gives the desired anti-Hermitian dynamics, and $\hat{E}_1^{\text{DC}}$ is the inevitable ``quantum jump'' term that comes from the requirement of trace completion. In the language of open quantum systems, $\hat{E}_1^{\text{DC}}$ is related to the Lindblad operators. 

This channel acts on the system qubit, and requires an ancillary qubit to implement, which extends the system Hilbert space to make the dynamics of the combined system unitary. This is achieved by the controlled-$y$ rotation gate, $CR_y(\varphi)$, where $\varphi=-2\arcsin{\sqrt{\gamma}}$, with the ancillary qubit as the target. In the computational basis, this is the matrix
\begin{equation}
    \hat{U}^{\text{DC}}=\mathbb{1}\otimes\hat{E}_0^{\text{DC}}-i\hat{\sigma}^y\otimes\hat{E}_1^{\text{DC}}=\begin{pmatrix}
        1 & 0 & 0 & 0 \\
        0 & \sqrt{1-\gamma} & 0 & \sqrt{\gamma} \\
        0 &  0 & 1 & 0 \\
        0 & -\sqrt{\gamma} & 0 & \sqrt{1-\gamma}
    \end{pmatrix},
\end{equation}
where the first term in the tensor products acts on the ancillary qubit and the second term acts on the system qubit. After one action of this unitary, with the ancillary qubit prepared in the $\ket{0}$ state, we get
\begin{equation}
    \hat{U}^{\text{DC}}\rho_{\text{tot}}\hat{U}^{\text{DC}\dagger}=\ket{0}\bra{0}\otimes\hat{E}_0^{\text{DC}}\rho\hat{E}_0^{\text{DC}\dagger}+\ket{1}\bra{1}\otimes\hat{E}_1^{\text{DC}}\rho\hat{E}_1^{\text{DC}\dagger}.
\end{equation}
Following this, a measurement is performed on the ancillary qubit. If it is found to be in the $\ket{0}$ state, the density matrix of the system has been acted on by the desired evolution according to $\hat{k}_j$ in Eq.~(\ref{eq:SingNHHam})
\begin{equation}
    \frac{\hat{E}_0^{\text{DC}}\rho\hat{E}_0^{\text{DC}\dagger}}{\Tr\left(\hat{E}_0^{\text{DC}}\rho\hat{E}_0^{\text{DC}\dagger}\right)}=\frac{e^{\Theta\hat{\sigma}_z\delta t}\rho e^{\Theta\hat{\sigma}_z\delta t}}{\Tr\left(e^{2\Theta\hat{\sigma}_z\delta t}\right)}.
\end{equation}
The probability of measuring the ancillary qubit in the $\ket{0}$ state is $p_0=\Tr\left(e^{2\Theta\hat{\sigma}_z\delta t}\rho\right)\approx 1-\frac{\gamma}{2}(1-r\cos\theta)$, where $(r,\theta,\phi)$ are the Bloch ball coordinates for a single qubit. If, on the other hand, the ancillary qubit is measure in the $\ket{1}$ state, the system is taken out of the physical subspace. After a single Trotter step, the ancillary qubit is reset to the $\ket{0}$ state and the quantum channel is applied again. For a large number of time-steps, this probability approaches $e^{-4\Theta\delta t}$; it becomes increasingly likely the system will leave the physical subspace. Nevertheless, we will show in this work that interesting properties of the system can be extracted at short time-scales on NISQ era devices.

%%%%%%%%%%%%%%%%%%%%%%%%%%%%%%%%%%%%%%%%%%%%%%%%%%%%%%%%%%%%%%%%%%%%%%%%%%%%%%%%
%%%%%%%%%%%%%%%%%%%%%%%%%%%%%%%%%%%%%%%%%%%%%%%%%%%%%%%%%%%%%%%%%%%%%%%%%%%%%%%%

\section{\label{sec:Model}Model}

Consider the 1D Ising model with an imaginary longitudinal magnetic field with open boundary conditions, given by the Hamiltonian
\begin{equation}\label{eq:HamAll}
    \hat{H}_{\text{all}} = -\lambda \sum_{i=1}^{N_s - 1} \hat{\sigma}^z_i \hat{\sigma}^z_{i + 1} - h_x \sum_{i = 1}^{N_s} \hat{\sigma}^x_i + i \Theta \sum_{i = 1}^{N_s} \hat{\sigma}^z_i.
\end{equation}
Under the Trotter approximation, the dynamics according to this Hamiltonian can be broken up into three pieces
\begin{equation}
    \hat{U}_{\text{full}} = e^{-i\hat{H}_{\text{all}}\delta t}\approx\hat{M}_{\text{U}}\hat{M}_{\text{NU}}\approx \hat{U}_{\text{NN}}\hat{U}_{\text{X}}\left(\hat{M}_{\text{NU}}\right),
\end{equation}
with
\begin{align}
    \hat{U}_{\text{NN}} &= e^{i \lambda \delta t \sum_{i=1}^{N_s - 1} \hat{\sigma}^z_i \hat{\sigma}^z_{i + 1} }, \\
    \hat{U}_{\text{X}} &= e^{i h_x \delta t \sum_{i = 1}^{N_s} \hat{\sigma}^x_i},
\end{align}
and where the last piece is the evolution according to the anti-Hermitian longitudinal magnetic field,
\begin{align}
    \hat{M}_{\text{NU}} = e^{\Theta \delta t \sum_{i = 1}^{N_s} \hat{\sigma}^z_i}.
\end{align}
The unitary terms are straightforward to implement using standard digital gates. The non-unitary terms require one ancilla qubit per system qubit to implement via the channel outlined in Sec.~\ref{sec:Theory}. Therefore, the full unitary implemented on the extended (system + ancillas) Hilbert space is
\begin{equation}
    \hat{U}_{\text{full}}\approx\left(\mathbb{1}\otimes\hat{U}_{\text{NN}}\right)\left(\mathbb{1}\otimes\hat{U}_{\text{X}}\right)\left(\hat{U}^{\text{DC}}\right)
\end{equation}
where $\hat{U}^{\text{DC}}$ acts on all system and ancillary qubits.

\begin{figure}[h!]
\center{
\includegraphics[width=0.24\textwidth]{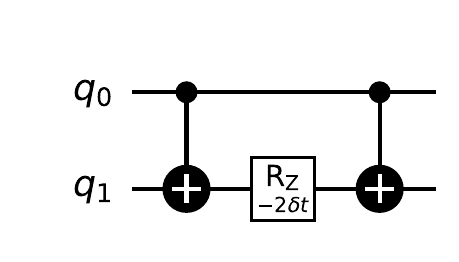}
\includegraphics[width=0.23\textwidth]{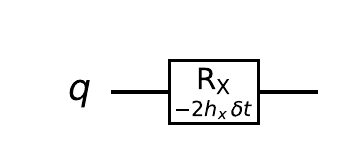}
\includegraphics[width=0.48\textwidth]{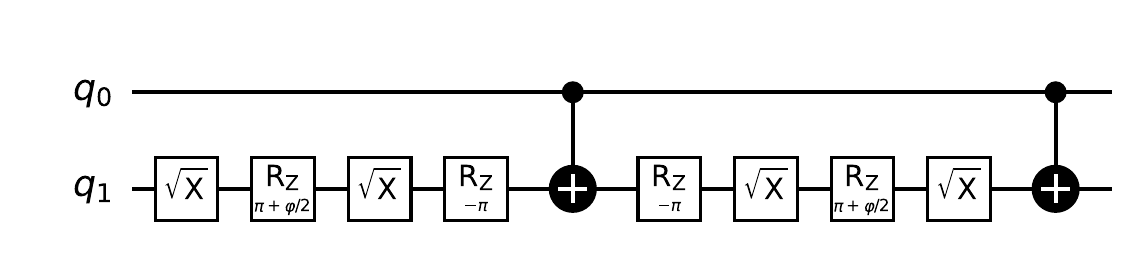}}
\caption{The decomposition into a basis gate set $[\text{CNOT, RZ, } \sqrt{\text{X}}, \text{X}]$ of time evolution according to (in reading order) the nearest-neighbour term, the transverse magnetic field term, and the unitary $U^{\text{DC}}$ acting on the extended system (with one ancillary qubit), corresponding to the imaginary longitudinal magnetic field.}
\label{fig:BlockCircs}
\end{figure}

One could in-principle use Quantum Shannon Decomposition (QSD) \cite{Shende_2006} to decompose the full time-evolution unitary $e^{-i\hat{H}\,t}$ for arbitrary $t$ for small system sizes. QSD decomposes any $N$-qubit unitary into a circuit that has at-most $\mathcal{O}(4^N)$ CNOT gates. However, finding the actual decomposition is a very difficult task for larger systems.  

The gate decomposition of these into a standard gate set is shown in Fig.~\ref{fig:BlockCircs}. These circuits form the building blocks of the full quantum circuit. The bottleneck in terms of hardware error rates comes from the two-qubit CNOT gates in NISQ-era devices. The scaling of the number of CNOT gates with the number of system qubits per Trotter step (first-order) is shown in column 2 of Table~\ref{tab:CNOTscaling}

\begin{table}[h!]
\centering
\begin{tabular}{||c | c | c||} 
 \hline
 Term & CNOTs (all) & CNOTs (alt)\\
 \hline\hline
 $\hat{U}_{\text{NN}}$ & $2(N-1)$ & $2(N-1)$ \\
 \hline
 $\hat{U}_{\text{X}}$ & $0$ & $0$ \\ 
 \hline
 $\hat{U}^{\text{DC}}$ & $2N$ & $N$ \\
 \hline
 \textbf{Total} & $2(2N-1)$ & $2(3/2N-1)$ \\
 \hline
\end{tabular}
\caption{Scaling of number of CNOTs with system size to implement one step of the dynamics of the Trotterized Hamiltonians in Eq.~(\ref{eq:HamAll}) (column 2) and Eq.~(\ref{eq:HamAlt})(column 3)}
\label{tab:CNOTscaling}
\end{table}

\begin{figure}
    \centering
    \includegraphics[width = 0.8\linewidth]{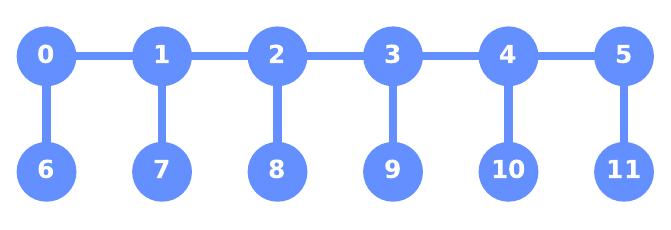}
    \caption{Ideal hardware topology for quantum simulation of the Hamiltonian in Eq.~(\ref{eq:HamAll})}
    \label{fig:CombTopo}
\end{figure}

We use IBM Quantum devices to simulate the Hamiltonian in this work. These devices have fixed topologies that are not ideal for simulating the Hamiltonian in Eq.~(\ref{eq:HamAll}) using the damping channel. The ideal topology for the Hamiltonian is a `comb' shown in Fig.~\ref{fig:CombTopo}. This mismatch necessitates the use of the expensive SWAP gates, requiring $3$ CNOT gates each. Of course, this extra cost is not present for very small system sizes like two or four sites, but starts to show up for even six system sites. To avoid this extra cost (and noise), we restrict ourselves to the case where the anti-Hermitian piece in Eq.~\ref{eq:HamAll} acts only on alternate sites of the 1D Ising chain
\begin{equation}\label{eq:HamAlt}
    \hat{H}_{\text{alt}} = -\lambda \sum_{i=1}^{N_s - 1} \hat{\sigma}^z_i \hat{\sigma}^z_{i + 1} - h_x \sum_{i = 1}^{N_s} \hat{\sigma}^x_i + i \Theta \sum_{i = 1}^{\lfloor N_s/2 \rfloor} \hat{\sigma}^z_{2i}.    
\end{equation}
This modified Hamiltonian also has the interesting features of non-Hermitian Hamiltonians we would like to observe through quantum computation. The scaling of the CNOT gate cost per (first-order) Trotter step for this modified Hamiltonian is shown in column 3 in Table~\ref{tab:CNOTscaling}.

\subsection{Exceptional points}

Non-Hermitian Hamiltonians have interesting features that we hope to see using the algorithm we have reviewed so far. In this subsection, we will briefly discuss what exceptional points are and why it is still possible to see them using noisy quantum devices.

Non-Hermitian Hamiltonians can have complex eigenvalues and non-orthonormal eigenvectors. In some regions of parameter space, the ground state of the Hamiltonian can be real. As one approaches an exceptional point from this region, the ground state and the first excited state approach each other and become degenerate on the exceptional point. Moreover, the corresponding eigenvectors merge and the rank of the Hamiltonian is reduced, allowing it to only be diagonalizable up to a Jordan block.

\begin{figure}[h!]
    \centering
    \includegraphics[width = 0.95\linewidth]{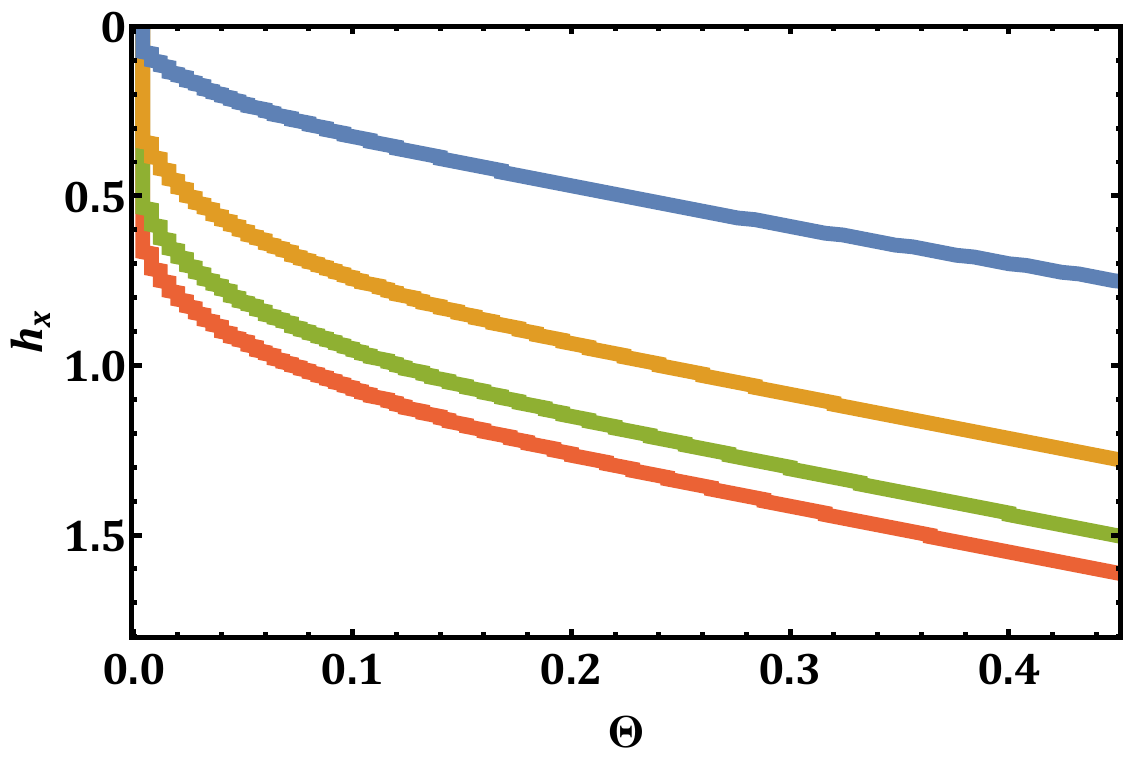}
    \caption{The line of exceptional points for the Hamiltonian in Eq.~(\ref{eq:HamAlt}) for (from top to bottom) 2, 4, 6, 8 site systems}
    \label{fig:ExcepPts}
\end{figure}

In Fig~\ref{fig:ExcepPts}, we show the line of exceptional points of the Hamiltonian in Eq.~(\ref{eq:HamAlt}) for 2, 4, 6, and 8 sites. $h_{\text{x}}$ and $\Theta$ are respectively the strength of the transverse magnetic field and the imaginary longitudinal magnetic field that acts on even sites. When $\Theta=0$, the usual Ising model is required. In the large system-size limit, the line of exceptional points deviates from the $\Theta=0$ line at $h_{\text{x}}=1$, which is the well-known phase transition of the 1D quantum Ising chain. In this limit, this line of exceptional points is called the Lee-Yang edge.

When you go beyond this line of exceptional points, the now degenerate ground state splits off into the complex plane as a complex conjugate pair. The state corresponding to the positive imaginary part of the eigenvalue becomes an attractor for the dynamics---time evolution according to the Hamiltonian of any state leads to this state eventually in this region of parameter space beyond the exceptional line. This is an appealing feature when simulating on noisy devices: Even if the noise kicks the state off the correct evolution, further time-evolution nudges the state back towards the attractor. So, the line of exceptional points should be a relatively robust feature visible even on noisy devices. 

The purity of the state of the system beyond the line of exceptional points will be close to unity, whereas the  for states below the line of exceptional points will be  mixed. An observable like the R\`{e}nyi entropy would be able to detect this feature. However, the purity of the state is not accessible without at least some level of tomography to find the off-diagonal elements of the reduced density matrix. This would not scale well for larger systems, so we have identified a simple diagonal observable that is sensitive to the line of exceptional points
\begin{equation}\label{eq:Observ}
    \hat{\mathcal{O}}=\frac{1}{N}\sum_{i}\hat{\sigma}_i^z,
\end{equation}
where $N$ is the number of system qubits, and the sum runs over all the system sites.

%%%%%%%%%%%%%%%%%%%%%%%%%%%%%%%%%%%%%%%%%%%%%%%%%%%%%%%%%%%%%%%%%%%%%%%%%%%%%%%%
%%%%%%%%%%%%%%%%%%%%%%%%%%%%%%%%%%%%%%%%%%%%%%%%%%%%%%%%%%%%%%%%%%%%%%%%%%%%%%%%

\section{\label{sec:Results}Results}

In this section, we briefly discuss the types of expected errors and how we mitigate them. Following that, we show the results from quantum computation on the \texttt{ibmq\textunderscore kolkata} device. 

Very broadly, there are two sources of errors in this system: The inevitable quantum jumps stemming from the fact that the system of interest has non-unitary dynamics and the hardware error. To deal with quantum jumps we only post-select the counts from the quantum computation where all the measurements of the ancillary qubits yielded $0$. Typically, this leads to `shot-noise', since the number of effective shots left over is a fraction of the total shots of the computation. This is remedied by taking a large number of shots per circuit, $100\,000$. Each circuit requires around $50$ seconds of quantum runtime usage.

The leading sources of hardware errors are measurement errors (we need a total of $N\left(\frac{1}{2}N_t +1\right)$ measurements, where $N_t$ is the number of Trotter steps), qubit decoherence, and CNOT gate errors. In this work we stick to a relatively light error mitigation strategy. We use the M3 (Matrix-free Measurement Mitigation) protocol~\cite{PRXQuantum.2.040326} and dynamical decoupling~\cite{PhysRevA.58.2733,ZANARDI199977,PhysRevA.59.4178,DUAN1999139,PhysRevLett.82.2417} to respectively tackle measurement errors and qubit decoherence. See \cite{charles2023simulating} for an analysis of different error mitigation strategies for quantum simulations. One could add heavier error mitigation strategies to deal with CNOT gate errors like zero-noise extrapolation, but the focus of this work is to demonstrate the resilience of exceptional points to hardware noise.

\subsection{M3 mitigation}

During the measurement stage of a quantum computation, there is a probability of readout assignment error. For a single qubit, this occurs when a $0(1)$ is incorrectly measured as a $1(0)$. The reason this happens is because qubit measurement is a process that takes a finite amount of time, allowing the qubit to interact with the environment. This type of error would affect the characterization of the state after several shots, and thus the observables being measured. Typically, this type of error is quantified by performing calibration measurements by state preparation and measurement to create a `confusion matrix' whose entries $p_{ij}$ are the probability of incorrectly measuring the $i^{\text{th}}$ bit-string as the $j^{\text{th}}$ bit-string. Then, one inverts this matrix and maps the noisy probability distribution to the ideal one. The full confusion matrix is $2^N\times 2^N$-dimensional, requiring an exponential amount of calibration measurements. For a large number of qubits, the matrix inversion process also becomes resource-intensive. The M3 protocol addresses both of these issues by restricting the confusion matrix only to the populated bitstrings in the noisy probability distribution and only considering errors between bitstrings that are a small Hamming distance away from each other. This makes the scaling, at most, linear in the number of calibration measurements and the matrix inversion procedure easier. For instance, for small enough systems, one could use the LU-decomposition to aid in matrix inversion.

\subsection{Dynamical Decoupling}

In current devices, the decoherence of a qubit via unwanted interaction with the environment is a significant source of noise. This type of noise is quantified using two decay constants- $T_1$ and $T_2$, which are respectively associated with the $\ket{1}$ state, and the $\ket{+}$ state decaying, thereby losing information about energy and phase. On IBM devices, both $T_{1,2}\sim\mathcal{O}(100)\mu s$. This type of noise is suppressed using dynamical decoupling. The broad idea is to hit idle qubits with a series of timed quick pulses (or single-qubit gates) that move the state around in the Bloch sphere in a way that cancels the action of the environment on average. The assumption here is that the pulses are quicker than the time scale in which the state of the environment changes. There are several implementations of the dynamical decoupling, with different choices of pulse sequences. See~\cite{ezzell2023dynamical} for a detailed study comparing various implementations.

\subsection{Hardware Results}

\begin{figure}
    \centering
    \includegraphics[width=0.95\linewidth]{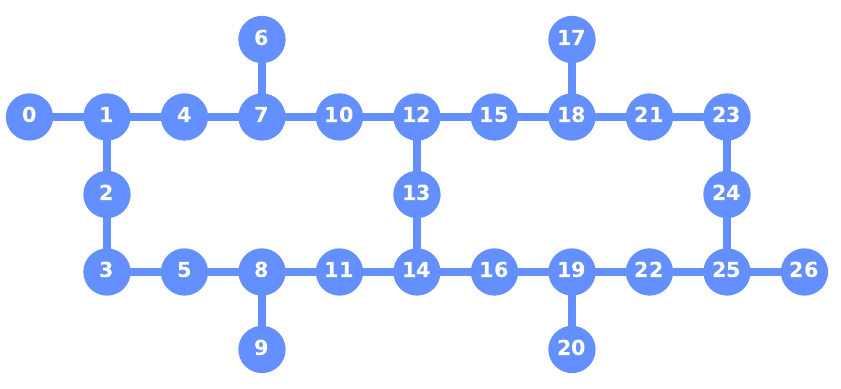}
    \caption{Hardware topology of \texttt{ibmq\textunderscore kolkata}}
    \label{fig:KolkataTopo}
\end{figure}

\begin{figure}[h!]
\center{
\includegraphics[width=0.40\textwidth]{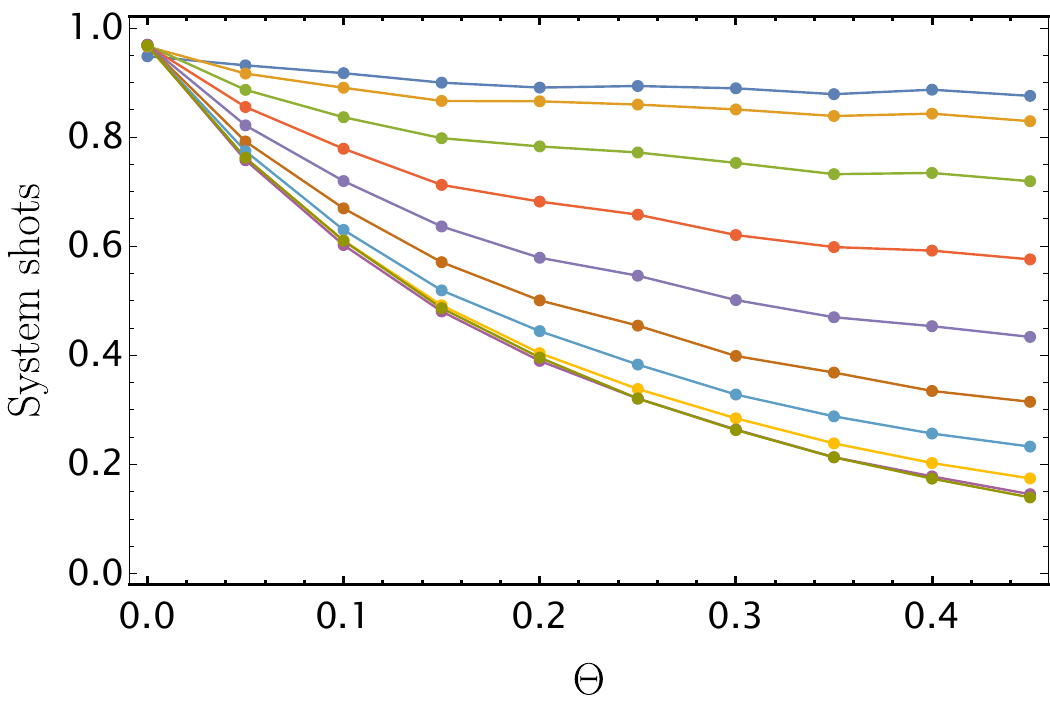}
\includegraphics[width=0.40\textwidth]{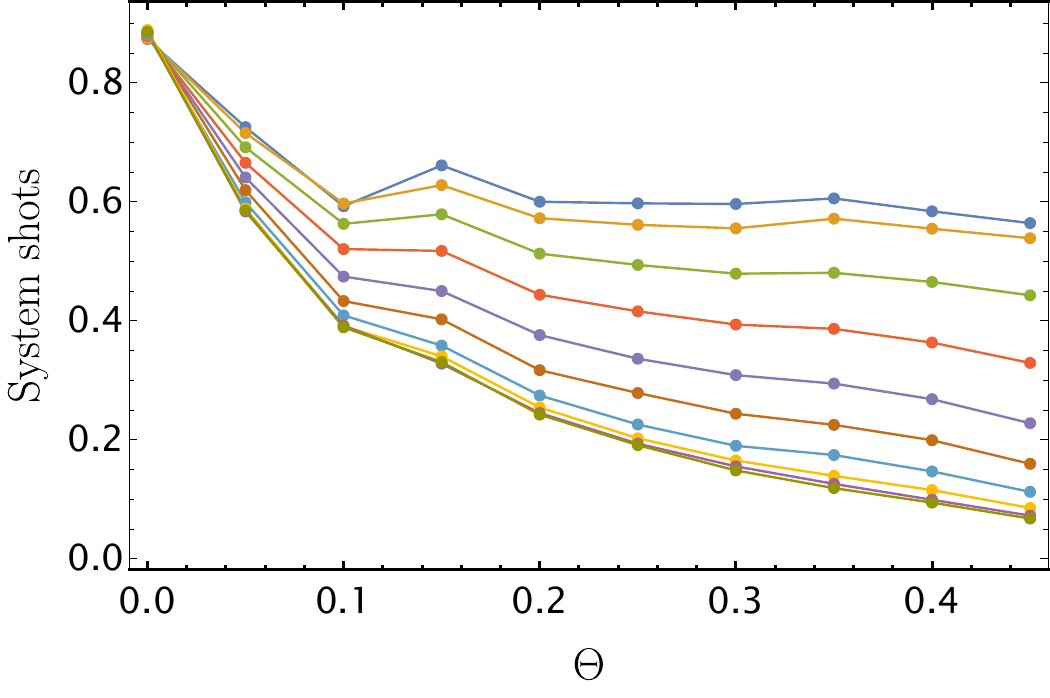}
\includegraphics[width=0.40\textwidth]{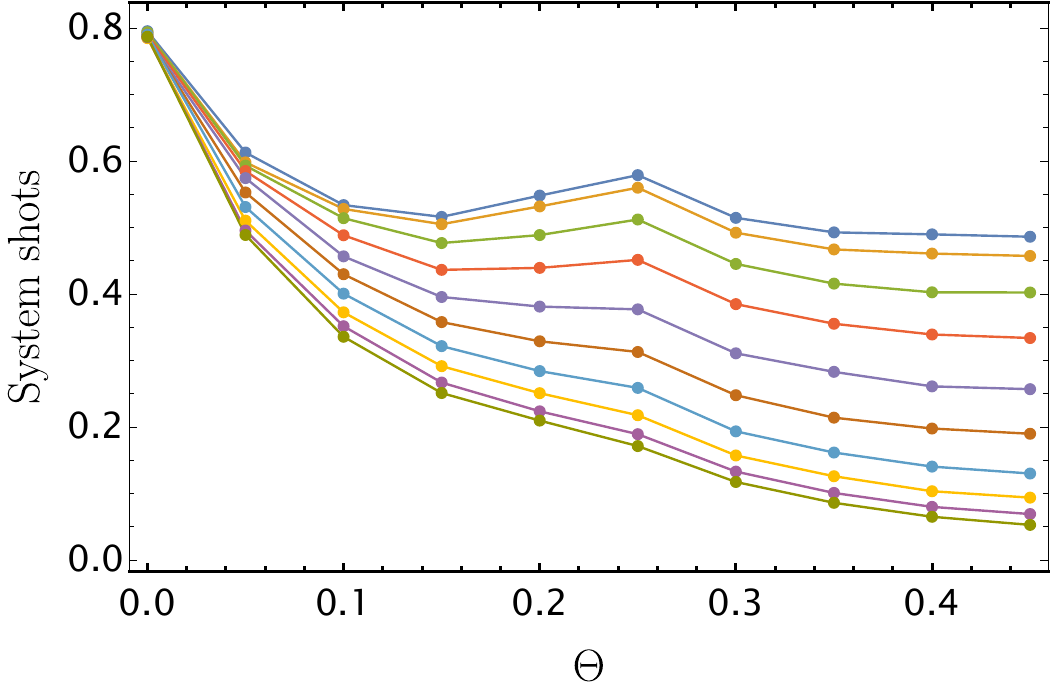}}
\includegraphics[width=0.40\textwidth]{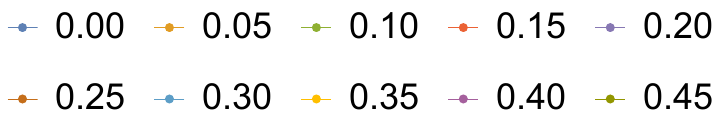}
\caption{The fraction of statistics with no quantum jumps for (from top to bottom) 2, 4, and 6 system sites after evolution by 7, 5, and 4 trotter steps respectively. The different colored lines correspond to values of $h_x$ shown in the legend}
\label{fig:SysStats}
\end{figure}

The topology of the hardware is shown in Fig.~\ref{fig:KolkataTopo}. We evolve the system according to the first-order Trotterized version of the Hamiltonian in Eq.~(\ref{eq:HamAlt}) for 2, 4, and 6 system qubits, with $dt = 0.3$. In Table~\ref{tab:QCdetails}, we show the particular qubits of the hardware we use, the number of Trotter steps, and the number of CNOT gates required for the circuits corresponding to various system sizes. In Fig~\ref{fig:SysStats} we show the fraction of post-selected statistics that do not have any quantum jumps for two, four, and six system sites. We provide the \texttt{qiskit} notebooks for building the circuits and running them on the hardware we used in \cite{GithubRepo}.

\begin{table}[h!]
\centering
\begin{tabular}{||c |c | c | c||} 
 \hline
 $N$ & Hardware qubits & $N_t$ & CNOTs \\
 \hline\hline
 $2$ & \texttt{[21, 23, (24)]} & $7$ & $28$ \\
 \hline
 $4$ & \texttt{[10, 12, 15, (18, 13, 17)]} & $5$ & $50$ \\
 \hline
 $6$ & \texttt{[10, 12, 15, 18, 21, 23,} & $4$ & $64$ \\
     & \texttt{(13, 17, 24)]} & & \\
 \hline 
\end{tabular}
\caption{Details of the quantum simulation on \texttt{ibmq\textunderscore kolkata} for various system sizes. The hardware qubits chosen as the ancillas are in parentheses.}
\label{tab:QCdetails}
\end{table}

\begin{figure}[h!]
\center{
\includegraphics[width=0.35\textwidth]{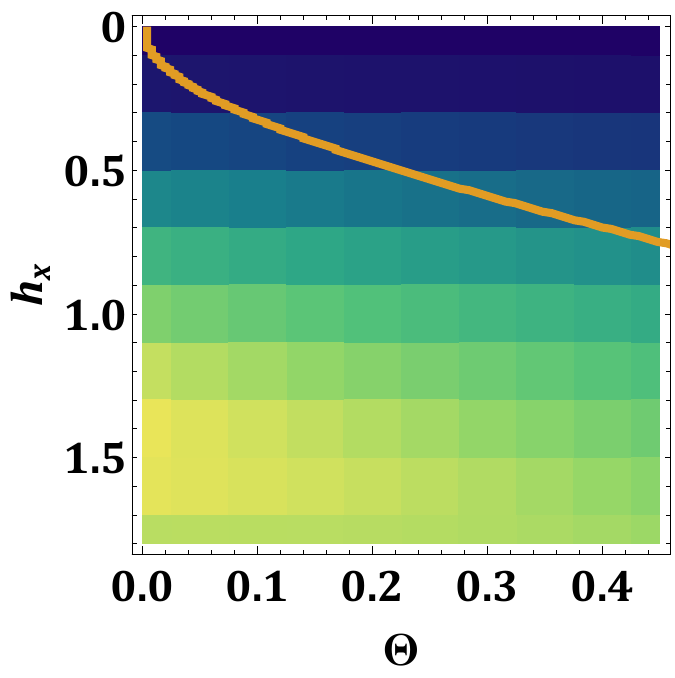}
\includegraphics[width=0.35\textwidth]{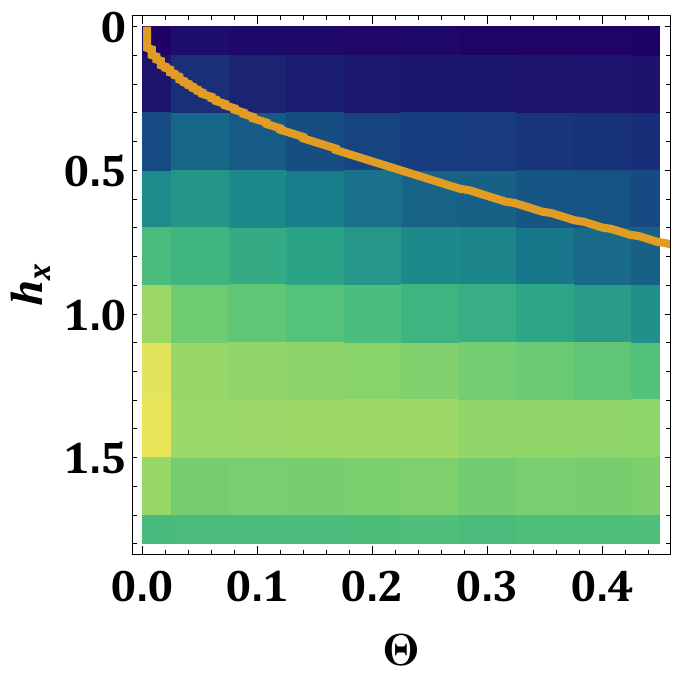}}
\includegraphics[width=0.35\textwidth]{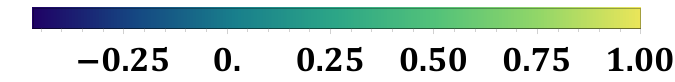}
\caption{$\langle\mathcal{O}\rangle$ from exact evolution, and the error-mitigated values from quantum computation for 2 system sites}
\label{fig:N2Plts}
\end{figure}

\begin{figure}[h!]
\center{
\includegraphics[width=0.35\textwidth]{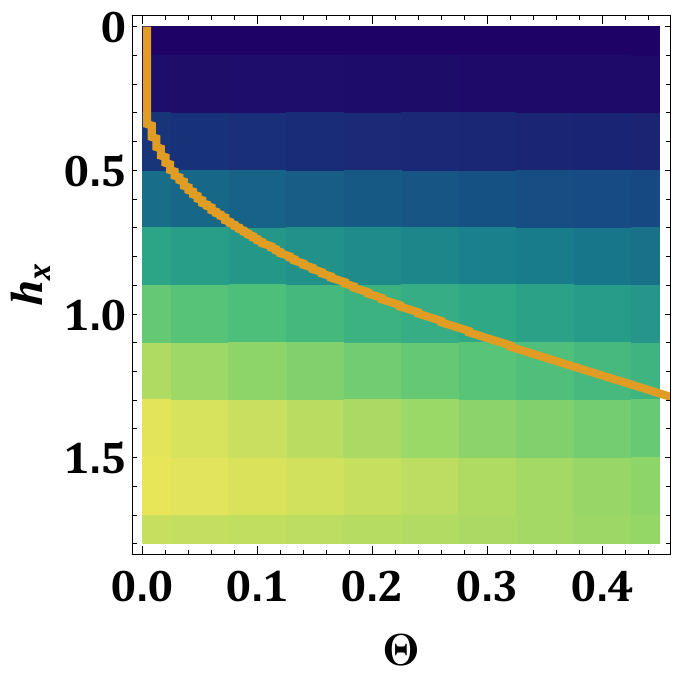}
\includegraphics[width=0.35\textwidth]{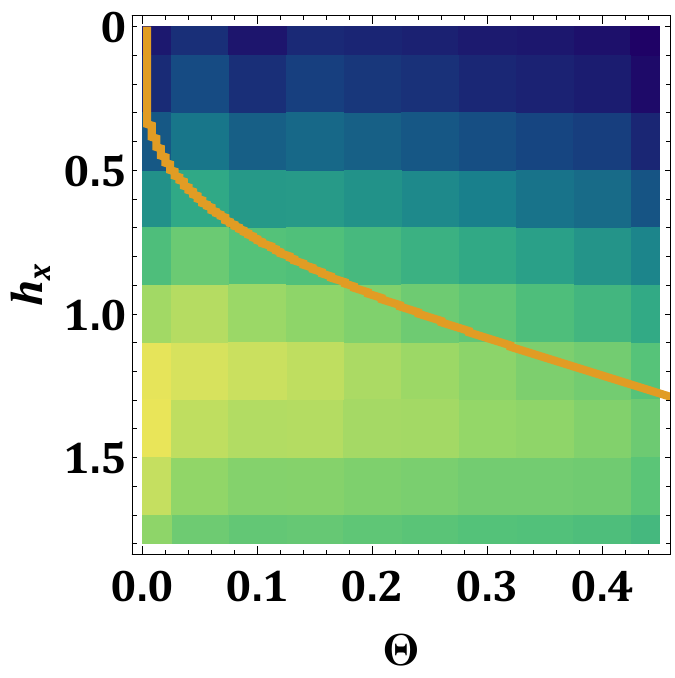}}
\includegraphics[width=0.35\textwidth]{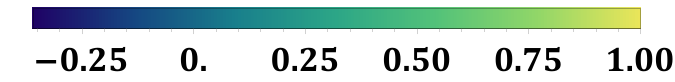}
\caption{$\langle\mathcal{O}\rangle$ from exact evolution, and the error-mitigated values from quantum computation for 4 system sites}
\label{fig:N4Plts}
\end{figure}

\begin{figure}[h!]
\center{
\includegraphics[width=0.35\textwidth]{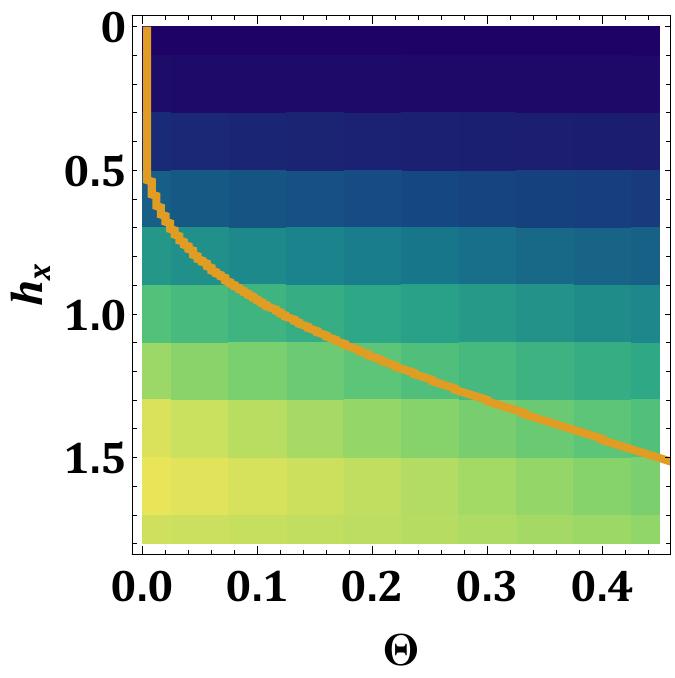}
\includegraphics[width=0.35\textwidth]{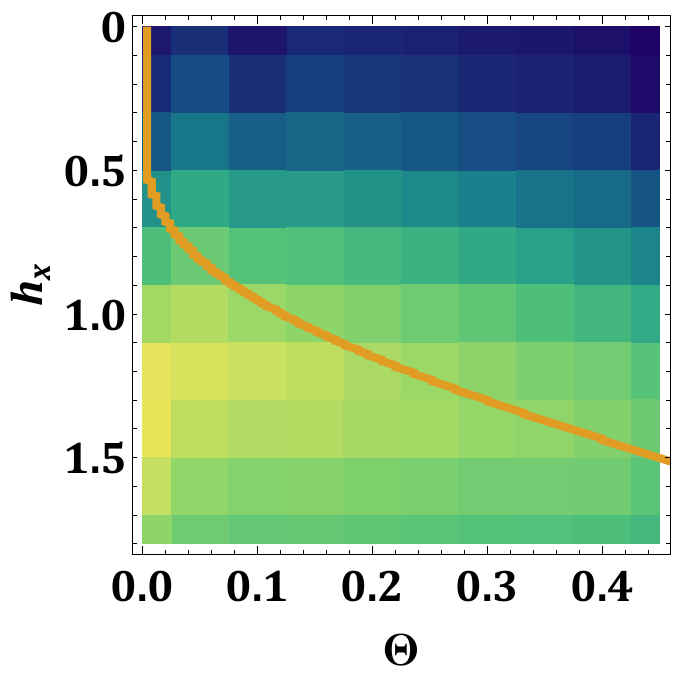}}
\includegraphics[width=0.35\textwidth]{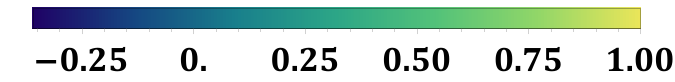}
\caption{$\langle\mathcal{O}\rangle$ from exact evolution, and the error-mitigated values from quantum computation for 6 system sites}
\label{fig:N6Plts}
\end{figure}

Following the evolution, we measure the diagonal observable in Eq.~(\ref{eq:Observ}) for $100$ equally spaced points in the $(\Theta,h_{\text{x}})$ parameter space. In Figs.~\ref{fig:N2Plts}, \ref{fig:N4Plts}, and \ref{fig:N6Plts}, respectively, we show the circuits for a Trotter step, the exact results, and quantum hardware computations on \texttt{ibmq\textunderscore kolkata} for two, four, and six system sites. The line of exceptional points is overlayed on top of both the exact and hardware simulation results. Comparing the quantum hardware results with the exact evolution, we see that the values of observables don't exactly match. This can be improved by using better error mitigation techniques. That being said, it is clear that the degree of sharpness of the exceptional point transition in the hardware computation is similar to the exact evolution. 

This is an example of what we are looking for---features of the model that are relatively robust to hardware noise. 
It is a promising sign that such features in non-Hermitian dynamics can be seen on NISQ-era devices.

\begin{figure}
    \centering
    \includegraphics[width=0.98\linewidth]{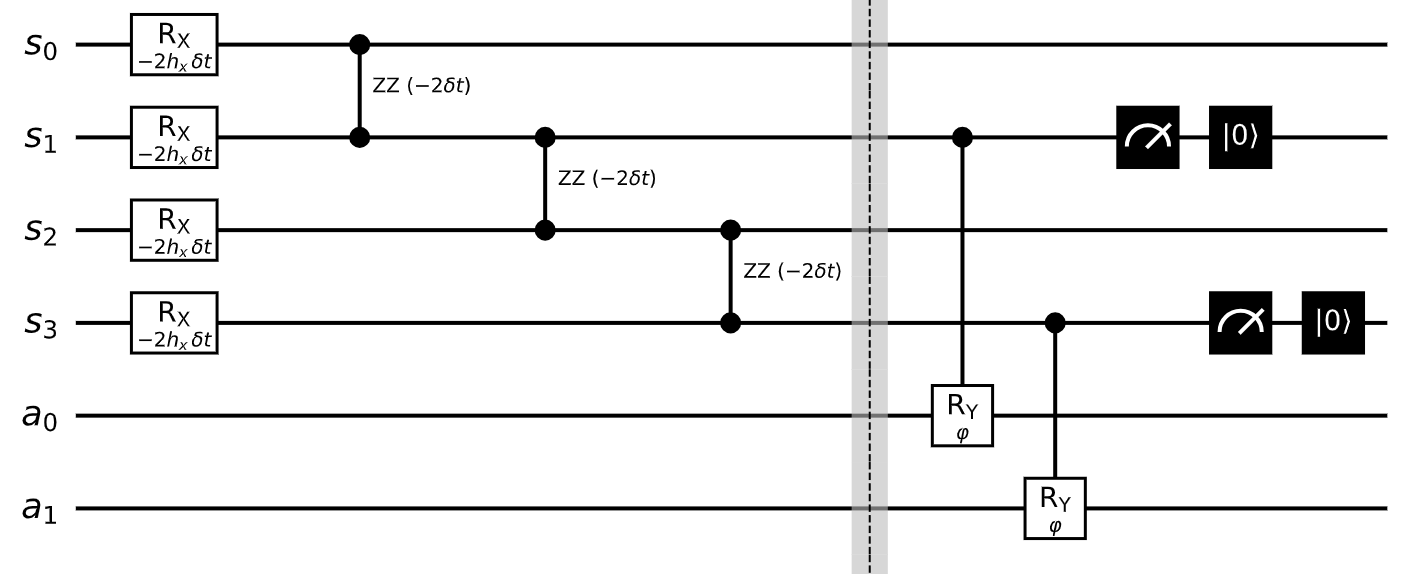}
    \caption{The circuit for one Trotter step for 4 system sites (and 2 ancillas).}
    \label{fig:4trot}
\end{figure}

%%%%%%%%%%%%%%%%%%%%%%%%%%%%%%%%%%%%%%%%%%%%%%%%%%%%%%%%%%%%%%%%%%%%%%%%%%%%%%%%
%%%%%%%%%%%%%%%%%%%%%%%%%%%%%%%%%%%%%%%%%%%%%%%%%%%%%%%%%%%%%%%%%%%%%%%%%%%%%%%%

\section{\label{sec:Conclusions}Conclusions and Outlook}
In this paper, we studied the hardware implementation of open quantum systems with non-Hermitian dynamics using the example of a 1D Ising model with an imaginary longitudinal magnetic field. We used a quantum channel approach developed in \cite{PhysRevA.104.052420} where the system of interest is embedded in a larger Hilbert space using ancillary qubits, with the enlarged system undergoing unitary evolution. The particular channel demonstrated in this work requires one ancillary qubit per system site that has the imaginary field acting on it. We study the dynamics under the first-order Trotter approximation, wherein the ancillary qubits are measured after each Trotter step. This process of measuring qubits during a computation can be expensive for a large number of Trotter steps since measurement of a qubit takes time. This creates a limitation due to the coherence times of contemporary qubits. The entanglement between the system and the ancillary qubits is set up in such a way that this measurement result reveals if the system has evolved according to the desired non-Hermitian Hamiltonian. The catch is that there is a small probability the system would undergo a quantum jump, putting the system in an undesired state. This can be combated using post-selection of results, which in-turn requires a large number of shots to ensure sufficient statistics in the physical subspace of interest.

There is another quantum channel, called the `Random Walk through time' also developed in \cite{PhysRevA.104.052420}, where the quantum jumps are controllable---they cause the system to Trotter evolve backwards in time instead. This could be useful from the standpoint of error mitigation in a way similar to the self-mitigation approach in~\cite{PhysRevD.106.074502}. Instead of building separate mitigation circuits, one could use some of the statistics of runs that had quantum jumps. Particularly, these would be the runs that evolved equally forward and backwards in time and could serve as a way to characterize the amount of noise in the circuit, provided the hardware errors are lesser than the Trotter error.

For some systems described by non-Hermitian effective Hamiltonians, there are interesting features in the complex parameter space.  Exceptional points, which give rise to an attractor evolution, are one such feature. We argue that as a result, exceptional points should be relatively resistant to hardware noise, and show for the example of a 1D Ising chain with an imaginary longitudinal magnetic field that it can be observed with some mild error mitigation strategies. The authors are hopeful that this would lead to the study of other interesting physical models using quantum computers without the need for heavy error mitigation protocols. 

%%%%%%%%%%%%%%%%%%%%%%%%%%%%%%%%%%%%%%%%%%%%%%%%%%%%%%%%%%%%%%%%%%%%%%%%%%%%%%%%
%%%%%%%%%%%%%%%%%%%%%%%%%%%%%%%%%%%%%%%%%%%%%%%%%%%%%%%%%%%%%%%%%%%%%%%%%%%%%%%%

\section{Acknowledgements}
Jay Hubisz and Bharath Sambasivam would like to thank the IBM-Q hub at Brookhaven National Laboratory for providing access to IBM quantum computers on a premium plan. This project would not have been possible without access to these valuable hardware resources. Jay Hubisz and
Bharath Sambasivam are supported in part by the U.S. Department of Energy (DOE), Office of Science, Office of High Energy Physics, under Award Number DE-SC0009998. Bharath Sambasivam would like to thank Dr. Muhammad Asaduzzaman and Prof. Jason Pollack for helpful discussions. Michael Hite was supported in part by the Department of Energy DE-SC0019139 and NSF award DMR-1747426.
This manuscript has been authored by Fermi Research Alliance, LLC under Contract No. DE-AC02-07CH11359 with the U.S. Department of Energy, Office of Science, Office of High Energy Physics. This work is supported by the Department of Energy through the Fermilab Theory QuantiSED program in the area of ``Intersections of QIS and Theoretical Particle Physics.''

%%%%%%%%%%%%%%%%%%%%%%%%%%%%%%%%%%%%%%%%%%%%%%%%%%%%%%%%%%%%%%%%%%%%%%%%%%%%%%%%
%%%%%%%%%%%%%%%%%%%%%%%%%%%%%%%%%%%%%%%%%%%%%%%%%%%%%%%%%%%%%%%%%%%%%%%%%%%%%%%%

%\clearpage
\bibliography{apssamp}% Produces the bibliography via BibTeX.

%apsrev4-2.bst 2019-01-14 (MD) hand-edited version of apsrev4-1.bst
%Control: key (0)
%Control: author (8) initials jnrlst
%Control: editor formatted (1) identically to author
%Control: production of article title (0) allowed
%Control: page (0) single
%Control: year (1) truncated
%Control: production of eprint (0) enabled
\providecommand{\noopsort}[1]{}\providecommand{\singleletter}[1]{#1}%
\begin{thebibliography}{48}%
\makeatletter
\providecommand \@ifxundefined [1]{%
 \@ifx{#1\undefined}
}%
\providecommand \@ifnum [1]{%
 \ifnum #1\expandafter \@firstoftwo
 \else \expandafter \@secondoftwo
 \fi
}%
\providecommand \@ifx [1]{%
 \ifx #1\expandafter \@firstoftwo
 \else \expandafter \@secondoftwo
 \fi
}%
\providecommand \natexlab [1]{#1}%
\providecommand \enquote  [1]{``#1''}%
\providecommand \bibnamefont  [1]{#1}%
\providecommand \bibfnamefont [1]{#1}%
\providecommand \citenamefont [1]{#1}%
\providecommand \href@noop [0]{\@secondoftwo}%
\providecommand \href [0]{\begingroup \@sanitize@url \@href}%
\providecommand \@href[1]{\@@startlink{#1}\@@href}%
\providecommand \@@href[1]{\endgroup#1\@@endlink}%
\providecommand \@sanitize@url [0]{\catcode `\\12\catcode `\$12\catcode `\&12\catcode `\#12\catcode `\^12\catcode `\_12\catcode `\%12\relax}%
\providecommand \@@startlink[1]{}%
\providecommand \@@endlink[0]{}%
\providecommand \url  [0]{\begingroup\@sanitize@url \@url }%
\providecommand \@url [1]{\endgroup\@href {#1}{\urlprefix }}%
\providecommand \urlprefix  [0]{URL }%
\providecommand \Eprint [0]{\href }%
\providecommand \doibase [0]{https://doi.org/}%
\providecommand \selectlanguage [0]{\@gobble}%
\providecommand \bibinfo  [0]{\@secondoftwo}%
\providecommand \bibfield  [0]{\@secondoftwo}%
\providecommand \translation [1]{[#1]}%
\providecommand \BibitemOpen [0]{}%
\providecommand \bibitemStop [0]{}%
\providecommand \bibitemNoStop [0]{.\EOS\space}%
\providecommand \EOS [0]{\spacefactor3000\relax}%
\providecommand \BibitemShut  [1]{\csname bibitem#1\endcsname}%
\let\auto@bib@innerbib\@empty
%</preamble>
\bibitem [{\citenamefont {Uzelac}\ \emph {et~al.}(1979)\citenamefont {Uzelac}, \citenamefont {Pfeuty},\ and\ \citenamefont {Jullien}}]{PhysRevLett.43.805}%
  \BibitemOpen
  \bibfield  {author} {\bibinfo {author} {\bibfnamefont {K.}~\bibnamefont {Uzelac}}, \bibinfo {author} {\bibfnamefont {P.}~\bibnamefont {Pfeuty}},\ and\ \bibinfo {author} {\bibfnamefont {R.}~\bibnamefont {Jullien}},\ }\bibfield  {title} {\bibinfo {title} {Yang-lee edge singularity from a real-space renormalization-group method},\ }\href {https://doi.org/10.1103/PhysRevLett.43.805} {\bibfield  {journal} {\bibinfo  {journal} {Phys. Rev. Lett.}\ }\textbf {\bibinfo {volume} {43}},\ \bibinfo {pages} {805} (\bibinfo {year} {1979})}\BibitemShut {NoStop}%
\bibitem [{\citenamefont {Itzykson}\ and\ \citenamefont {Drouffe}(1989)}]{itzykson_drouffe_1989}%
  \BibitemOpen
  \bibfield  {author} {\bibinfo {author} {\bibfnamefont {C.}~\bibnamefont {Itzykson}}\ and\ \bibinfo {author} {\bibfnamefont {J.-M.}\ \bibnamefont {Drouffe}},\ }\href {https://doi.org/10.1017/CBO9780511622779} {\emph {\bibinfo {title} {Statistical Field Theory}}},\ \bibinfo {series} {Cambridge Monographs on Mathematical Physics}, Vol.~\bibinfo {volume} {1}\ (\bibinfo  {publisher} {Cambridge University Press},\ \bibinfo {year} {1989})\BibitemShut {NoStop}%
\bibitem [{\citenamefont {Uzelac}\ \emph {et~al.}(1980)\citenamefont {Uzelac}, \citenamefont {Pfeuty},\ and\ \citenamefont {Jullien}}]{UZELAC19801011}%
  \BibitemOpen
  \bibfield  {author} {\bibinfo {author} {\bibfnamefont {K.}~\bibnamefont {Uzelac}}, \bibinfo {author} {\bibfnamefont {P.}~\bibnamefont {Pfeuty}},\ and\ \bibinfo {author} {\bibfnamefont {R.}~\bibnamefont {Jullien}},\ }\bibfield  {title} {\bibinfo {title} {1d transverse ising model in a longitudinal real or complex field},\ }\href {https://doi.org/https://doi.org/10.1016/0304-8853(80)90864-1} {\bibfield  {journal} {\bibinfo  {journal} {Journal of Magnetism and Magnetic Materials}\ }\textbf {\bibinfo {volume} {15-18}},\ \bibinfo {pages} {1011} (\bibinfo {year} {1980})}\BibitemShut {NoStop}%
\bibitem [{\citenamefont {von Gehlen}(1991)}]{GvonGehlen_1991}%
  \BibitemOpen
  \bibfield  {author} {\bibinfo {author} {\bibfnamefont {G.}~\bibnamefont {von Gehlen}},\ }\bibfield  {title} {\bibinfo {title} {Critical and off-critical conformal analysis of the ising quantum chain in an imaginary field},\ }\href {https://doi.org/10.1088/0305-4470/24/22/021} {\bibfield  {journal} {\bibinfo  {journal} {Journal of Physics A: Mathematical and General}\ }\textbf {\bibinfo {volume} {24}},\ \bibinfo {pages} {5371} (\bibinfo {year} {1991})}\BibitemShut {NoStop}%
\bibitem [{\citenamefont {Bender}\ and\ \citenamefont {Boettcher}(1998)}]{PhysRevLett.80.5243}%
  \BibitemOpen
  \bibfield  {author} {\bibinfo {author} {\bibfnamefont {C.~M.}\ \bibnamefont {Bender}}\ and\ \bibinfo {author} {\bibfnamefont {S.}~\bibnamefont {Boettcher}},\ }\bibfield  {title} {\bibinfo {title} {Real spectra in non-hermitian hamiltonians having pt symmetry},\ }\href {https://doi.org/10.1103/PhysRevLett.80.5243} {\bibfield  {journal} {\bibinfo  {journal} {Phys. Rev. Lett.}\ }\textbf {\bibinfo {volume} {80}},\ \bibinfo {pages} {5243} (\bibinfo {year} {1998})}\BibitemShut {NoStop}%
\bibitem [{\citenamefont {Bender}\ \emph {et~al.}(1999)\citenamefont {Bender}, \citenamefont {Boettcher},\ and\ \citenamefont {Meisinger}}]{10.1063/1.532860}%
  \BibitemOpen
  \bibfield  {author} {\bibinfo {author} {\bibfnamefont {C.~M.}\ \bibnamefont {Bender}}, \bibinfo {author} {\bibfnamefont {S.}~\bibnamefont {Boettcher}},\ and\ \bibinfo {author} {\bibfnamefont {P.~N.}\ \bibnamefont {Meisinger}},\ }\bibfield  {title} {\bibinfo {title} {{PT-symmetric quantum mechanics}},\ }\href {https://doi.org/10.1063/1.532860} {\bibfield  {journal} {\bibinfo  {journal} {Journal of Mathematical Physics}\ }\textbf {\bibinfo {volume} {40}},\ \bibinfo {pages} {2201} (\bibinfo {year} {1999})},\ \Eprint {https://arxiv.org/abs/https://pubs.aip.org/aip/jmp/article-pdf/40/5/2201/8169295/2201\_1\_online.pdf} {https://pubs.aip.org/aip/jmp/article-pdf/40/5/2201/8169295/2201\_1\_online.pdf} \BibitemShut {NoStop}%
\bibitem [{\citenamefont {Guo}\ \emph {et~al.}(2009)\citenamefont {Guo}, \citenamefont {Salamo}, \citenamefont {Duchesne}, \citenamefont {Morandotti}, \citenamefont {Volatier-Ravat}, \citenamefont {Aimez}, \citenamefont {Siviloglou},\ and\ \citenamefont {Christodoulides}}]{Guo:2009yqd}%
  \BibitemOpen
  \bibfield  {author} {\bibinfo {author} {\bibfnamefont {A.}~\bibnamefont {Guo}}, \bibinfo {author} {\bibfnamefont {G.~J.}\ \bibnamefont {Salamo}}, \bibinfo {author} {\bibfnamefont {D.}~\bibnamefont {Duchesne}}, \bibinfo {author} {\bibfnamefont {R.}~\bibnamefont {Morandotti}}, \bibinfo {author} {\bibfnamefont {M.}~\bibnamefont {Volatier-Ravat}}, \bibinfo {author} {\bibfnamefont {V.}~\bibnamefont {Aimez}}, \bibinfo {author} {\bibfnamefont {G.~A.}\ \bibnamefont {Siviloglou}},\ and\ \bibinfo {author} {\bibfnamefont {D.~N.}\ \bibnamefont {Christodoulides}},\ }\bibfield  {title} {\bibinfo {title} {{Observation of PT-Symmetry Breaking in Complex Optical Potentials}},\ }\href {https://doi.org/10.1103/PhysRevLett.103.093902} {\bibfield  {journal} {\bibinfo  {journal} {Phys. Rev. Lett.}\ }\textbf {\bibinfo {volume} {103}},\ \bibinfo {pages} {093902} (\bibinfo {year} {2009})}\BibitemShut {NoStop}%
\bibitem [{\citenamefont {Hahn}(1950)}]{PhysRev.80.580}%
  \BibitemOpen
  \bibfield  {author} {\bibinfo {author} {\bibfnamefont {E.~L.}\ \bibnamefont {Hahn}},\ }\bibfield  {title} {\bibinfo {title} {Spin echoes},\ }\href {https://doi.org/10.1103/PhysRev.80.580} {\bibfield  {journal} {\bibinfo  {journal} {Phys. Rev.}\ }\textbf {\bibinfo {volume} {80}},\ \bibinfo {pages} {580} (\bibinfo {year} {1950})}\BibitemShut {NoStop}%
\bibitem [{\citenamefont {Ezzell}\ \emph {et~al.}(2023)\citenamefont {Ezzell}, \citenamefont {Pokharel}, \citenamefont {Tewala}, \citenamefont {Quiroz},\ and\ \citenamefont {Lidar}}]{ezzell2023dynamical}%
  \BibitemOpen
  \bibfield  {author} {\bibinfo {author} {\bibfnamefont {N.}~\bibnamefont {Ezzell}}, \bibinfo {author} {\bibfnamefont {B.}~\bibnamefont {Pokharel}}, \bibinfo {author} {\bibfnamefont {L.}~\bibnamefont {Tewala}}, \bibinfo {author} {\bibfnamefont {G.}~\bibnamefont {Quiroz}},\ and\ \bibinfo {author} {\bibfnamefont {D.~A.}\ \bibnamefont {Lidar}},\ }\href@noop {} {\bibinfo {title} {Dynamical decoupling for superconducting qubits: a performance survey}} (\bibinfo {year} {2023}),\ \Eprint {https://arxiv.org/abs/2207.03670} {arXiv:2207.03670 [quant-ph]} \BibitemShut {NoStop}%
\bibitem [{\citenamefont {Bravyi}\ \emph {et~al.}(2021)\citenamefont {Bravyi}, \citenamefont {Sheldon}, \citenamefont {Kandala}, \citenamefont {Mckay},\ and\ \citenamefont {Gambetta}}]{Bravyi_2021}%
  \BibitemOpen
  \bibfield  {author} {\bibinfo {author} {\bibfnamefont {S.}~\bibnamefont {Bravyi}}, \bibinfo {author} {\bibfnamefont {S.}~\bibnamefont {Sheldon}}, \bibinfo {author} {\bibfnamefont {A.}~\bibnamefont {Kandala}}, \bibinfo {author} {\bibfnamefont {D.~C.}\ \bibnamefont {Mckay}},\ and\ \bibinfo {author} {\bibfnamefont {J.~M.}\ \bibnamefont {Gambetta}},\ }\bibfield  {title} {\bibinfo {title} {Mitigating measurement errors in multiqubit experiments},\ }\bibfield  {journal} {\bibinfo  {journal} {Physical Review A}\ }\textbf {\bibinfo {volume} {103}},\ \href {https://doi.org/10.1103/physreva.103.042605} {10.1103/physreva.103.042605} (\bibinfo {year} {2021})\BibitemShut {NoStop}%
\bibitem [{\citenamefont {Nation}\ \emph {et~al.}(2021)\citenamefont {Nation}, \citenamefont {Kang}, \citenamefont {Sundaresan},\ and\ \citenamefont {Gambetta}}]{PRXQuantum.2.040326}%
  \BibitemOpen
  \bibfield  {author} {\bibinfo {author} {\bibfnamefont {P.~D.}\ \bibnamefont {Nation}}, \bibinfo {author} {\bibfnamefont {H.}~\bibnamefont {Kang}}, \bibinfo {author} {\bibfnamefont {N.}~\bibnamefont {Sundaresan}},\ and\ \bibinfo {author} {\bibfnamefont {J.~M.}\ \bibnamefont {Gambetta}},\ }\bibfield  {title} {\bibinfo {title} {Scalable mitigation of measurement errors on quantum computers},\ }\href {https://doi.org/10.1103/PRXQuantum.2.040326} {\bibfield  {journal} {\bibinfo  {journal} {PRX Quantum}\ }\textbf {\bibinfo {volume} {2}},\ \bibinfo {pages} {040326} (\bibinfo {year} {2021})}\BibitemShut {NoStop}%
\bibitem [{\citenamefont {Nachman}\ \emph {et~al.}(2020)\citenamefont {Nachman}, \citenamefont {Urbanek}, \citenamefont {de~Jong},\ and\ \citenamefont {Bauer}}]{nachman2020unfolding}%
  \BibitemOpen
  \bibfield  {author} {\bibinfo {author} {\bibfnamefont {B.}~\bibnamefont {Nachman}}, \bibinfo {author} {\bibfnamefont {M.}~\bibnamefont {Urbanek}}, \bibinfo {author} {\bibfnamefont {W.~A.}\ \bibnamefont {de~Jong}},\ and\ \bibinfo {author} {\bibfnamefont {C.~W.}\ \bibnamefont {Bauer}},\ }\href@noop {} {\bibinfo {title} {Unfolding quantum computer readout noise}} (\bibinfo {year} {2020}),\ \Eprint {https://arxiv.org/abs/1910.01969} {arXiv:1910.01969 [quant-ph]} \BibitemShut {NoStop}%
\bibitem [{\citenamefont {Bennett}\ \emph {et~al.}(1996{\natexlab{a}})\citenamefont {Bennett}, \citenamefont {DiVincenzo}, \citenamefont {Smolin},\ and\ \citenamefont {Wootters}}]{PhysRevA.54.3824}%
  \BibitemOpen
  \bibfield  {author} {\bibinfo {author} {\bibfnamefont {C.~H.}\ \bibnamefont {Bennett}}, \bibinfo {author} {\bibfnamefont {D.~P.}\ \bibnamefont {DiVincenzo}}, \bibinfo {author} {\bibfnamefont {J.~A.}\ \bibnamefont {Smolin}},\ and\ \bibinfo {author} {\bibfnamefont {W.~K.}\ \bibnamefont {Wootters}},\ }\bibfield  {title} {\bibinfo {title} {Mixed-state entanglement and quantum error correction},\ }\href {https://doi.org/10.1103/PhysRevA.54.3824} {\bibfield  {journal} {\bibinfo  {journal} {Phys. Rev. A}\ }\textbf {\bibinfo {volume} {54}},\ \bibinfo {pages} {3824} (\bibinfo {year} {1996}{\natexlab{a}})}\BibitemShut {NoStop}%
\bibitem [{\citenamefont {Bennett}\ \emph {et~al.}(1996{\natexlab{b}})\citenamefont {Bennett}, \citenamefont {Brassard}, \citenamefont {Popescu}, \citenamefont {Schumacher}, \citenamefont {Smolin},\ and\ \citenamefont {Wootters}}]{PhysRevLett.76.722}%
  \BibitemOpen
  \bibfield  {author} {\bibinfo {author} {\bibfnamefont {C.~H.}\ \bibnamefont {Bennett}}, \bibinfo {author} {\bibfnamefont {G.}~\bibnamefont {Brassard}}, \bibinfo {author} {\bibfnamefont {S.}~\bibnamefont {Popescu}}, \bibinfo {author} {\bibfnamefont {B.}~\bibnamefont {Schumacher}}, \bibinfo {author} {\bibfnamefont {J.~A.}\ \bibnamefont {Smolin}},\ and\ \bibinfo {author} {\bibfnamefont {W.~K.}\ \bibnamefont {Wootters}},\ }\bibfield  {title} {\bibinfo {title} {Purification of noisy entanglement and faithful teleportation via noisy channels},\ }\href {https://doi.org/10.1103/PhysRevLett.76.722} {\bibfield  {journal} {\bibinfo  {journal} {Phys. Rev. Lett.}\ }\textbf {\bibinfo {volume} {76}},\ \bibinfo {pages} {722} (\bibinfo {year} {1996}{\natexlab{b}})}\BibitemShut {NoStop}%
\bibitem [{\citenamefont {Temme}\ \emph {et~al.}(2017)\citenamefont {Temme}, \citenamefont {Bravyi},\ and\ \citenamefont {Gambetta}}]{PhysRevLett.119.180509}%
  \BibitemOpen
  \bibfield  {author} {\bibinfo {author} {\bibfnamefont {K.}~\bibnamefont {Temme}}, \bibinfo {author} {\bibfnamefont {S.}~\bibnamefont {Bravyi}},\ and\ \bibinfo {author} {\bibfnamefont {J.~M.}\ \bibnamefont {Gambetta}},\ }\bibfield  {title} {\bibinfo {title} {Error mitigation for short-depth quantum circuits},\ }\href {https://doi.org/10.1103/PhysRevLett.119.180509} {\bibfield  {journal} {\bibinfo  {journal} {Phys. Rev. Lett.}\ }\textbf {\bibinfo {volume} {119}},\ \bibinfo {pages} {180509} (\bibinfo {year} {2017})}\BibitemShut {NoStop}%
\bibitem [{\citenamefont {Li}\ and\ \citenamefont {Benjamin}(2017)}]{PhysRevX.7.021050}%
  \BibitemOpen
  \bibfield  {author} {\bibinfo {author} {\bibfnamefont {Y.}~\bibnamefont {Li}}\ and\ \bibinfo {author} {\bibfnamefont {S.~C.}\ \bibnamefont {Benjamin}},\ }\bibfield  {title} {\bibinfo {title} {Efficient variational quantum simulator incorporating active error minimization},\ }\href {https://doi.org/10.1103/PhysRevX.7.021050} {\bibfield  {journal} {\bibinfo  {journal} {Phys. Rev. X}\ }\textbf {\bibinfo {volume} {7}},\ \bibinfo {pages} {021050} (\bibinfo {year} {2017})}\BibitemShut {NoStop}%
\bibitem [{\citenamefont {Kandala}\ \emph {et~al.}(2019)\citenamefont {Kandala}, \citenamefont {Temme}, \citenamefont {C{\'{o}}rcoles}, \citenamefont {Mezzacapo}, \citenamefont {Chow},\ and\ \citenamefont {Gambetta}}]{Kandala2019}%
  \BibitemOpen
  \bibfield  {author} {\bibinfo {author} {\bibfnamefont {A.}~\bibnamefont {Kandala}}, \bibinfo {author} {\bibfnamefont {K.}~\bibnamefont {Temme}}, \bibinfo {author} {\bibfnamefont {A.~D.}\ \bibnamefont {C{\'{o}}rcoles}}, \bibinfo {author} {\bibfnamefont {A.}~\bibnamefont {Mezzacapo}}, \bibinfo {author} {\bibfnamefont {J.~M.}\ \bibnamefont {Chow}},\ and\ \bibinfo {author} {\bibfnamefont {J.~M.}\ \bibnamefont {Gambetta}},\ }\bibfield  {title} {\bibinfo {title} {Error mitigation extends the computational reach of a noisy quantum processor},\ }\href {https://doi.org/10.1038/s41586-019-1040-7} {\bibfield  {journal} {\bibinfo  {journal} {Nature}\ }\textbf {\bibinfo {volume} {567}},\ \bibinfo {pages} {491} (\bibinfo {year} {2019})}\BibitemShut {NoStop}%
\bibitem [{\citenamefont {Urbanek}\ \emph {et~al.}(2021)\citenamefont {Urbanek}, \citenamefont {Nachman}, \citenamefont {Pascuzzi}, \citenamefont {He}, \citenamefont {Bauer},\ and\ \citenamefont {de~Jong}}]{Urbanek_2021}%
  \BibitemOpen
  \bibfield  {author} {\bibinfo {author} {\bibfnamefont {M.}~\bibnamefont {Urbanek}}, \bibinfo {author} {\bibfnamefont {B.}~\bibnamefont {Nachman}}, \bibinfo {author} {\bibfnamefont {V.~R.}\ \bibnamefont {Pascuzzi}}, \bibinfo {author} {\bibfnamefont {A.}~\bibnamefont {He}}, \bibinfo {author} {\bibfnamefont {C.~W.}\ \bibnamefont {Bauer}},\ and\ \bibinfo {author} {\bibfnamefont {W.~A.}\ \bibnamefont {de~Jong}},\ }\bibfield  {title} {\bibinfo {title} {Mitigating depolarizing noise on quantum computers with noise-estimation circuits},\ }\bibfield  {journal} {\bibinfo  {journal} {Physical Review Letters}\ }\textbf {\bibinfo {volume} {127}},\ \href {https://doi.org/10.1103/physrevlett.127.270502} {10.1103/physrevlett.127.270502} (\bibinfo {year} {2021})\BibitemShut {NoStop}%
\bibitem [{OBr(2023)}]{OBrien2023}%
  \BibitemOpen
  \bibfield  {title} {\bibinfo {title} {Purification-based quantum error mitigation of pair-correlated electron simulations},\ }\bibfield  {journal} {\bibinfo  {journal} {Nature Physics}\ }\href {https://doi.org/10.1038/s41567-023-02240-y} {10.1038/s41567-023-02240-y} (\bibinfo {year} {2023})\BibitemShut {NoStop}%
\bibitem [{\citenamefont {Strikis}\ \emph {et~al.}(2021)\citenamefont {Strikis}, \citenamefont {Qin}, \citenamefont {Chen}, \citenamefont {Benjamin},\ and\ \citenamefont {Li}}]{Strikis_2021}%
  \BibitemOpen
  \bibfield  {author} {\bibinfo {author} {\bibfnamefont {A.}~\bibnamefont {Strikis}}, \bibinfo {author} {\bibfnamefont {D.}~\bibnamefont {Qin}}, \bibinfo {author} {\bibfnamefont {Y.}~\bibnamefont {Chen}}, \bibinfo {author} {\bibfnamefont {S.~C.}\ \bibnamefont {Benjamin}},\ and\ \bibinfo {author} {\bibfnamefont {Y.}~\bibnamefont {Li}},\ }\bibfield  {title} {\bibinfo {title} {Learning-based quantum error mitigation},\ }\bibfield  {journal} {\bibinfo  {journal} {{PRX} Quantum}\ }\textbf {\bibinfo {volume} {2}},\ \href {https://doi.org/10.1103/prxquantum.2.040330} {10.1103/prxquantum.2.040330} (\bibinfo {year} {2021})\BibitemShut {NoStop}%
\bibitem [{\citenamefont {Qin}\ \emph {et~al.}(2023)\citenamefont {Qin}, \citenamefont {Chen},\ and\ \citenamefont {Li}}]{Qin2023}%
  \BibitemOpen
  \bibfield  {author} {\bibinfo {author} {\bibfnamefont {D.}~\bibnamefont {Qin}}, \bibinfo {author} {\bibfnamefont {Y.}~\bibnamefont {Chen}},\ and\ \bibinfo {author} {\bibfnamefont {Y.}~\bibnamefont {Li}},\ }\bibfield  {title} {\bibinfo {title} {Error statistics and scalability of quantum error mitigation formulas},\ }\bibfield  {journal} {\bibinfo  {journal} {npj Quantum Information}\ }\textbf {\bibinfo {volume} {9}},\ \href {https://doi.org/10.1038/s41534-023-00707-7} {10.1038/s41534-023-00707-7} (\bibinfo {year} {2023})\BibitemShut {NoStop}%
\bibitem [{\citenamefont {{Qiskit contributors}}(2023)}]{Qiskit}%
  \BibitemOpen
  \bibfield  {author} {\bibinfo {author} {\bibnamefont {{Qiskit contributors}}},\ }\href {https://doi.org/10.5281/zenodo.2573505} {\bibinfo {title} {Qiskit: An open-source framework for quantum computing}} (\bibinfo {year} {2023})\BibitemShut {NoStop}%
\bibitem [{\citenamefont {Su}\ and\ \citenamefont {Li}(2020)}]{Su_2020}%
  \BibitemOpen
  \bibfield  {author} {\bibinfo {author} {\bibfnamefont {H.-Y.}\ \bibnamefont {Su}}\ and\ \bibinfo {author} {\bibfnamefont {Y.}~\bibnamefont {Li}},\ }\bibfield  {title} {\bibinfo {title} {Quantum algorithm for the simulation of open-system dynamics and thermalization},\ }\bibfield  {journal} {\bibinfo  {journal} {Physical Review A}\ }\textbf {\bibinfo {volume} {101}},\ \href {https://doi.org/10.1103/physreva.101.012328} {10.1103/physreva.101.012328} (\bibinfo {year} {2020})\BibitemShut {NoStop}%
\bibitem [{\citenamefont {Guimarães}\ \emph {et~al.}(2023)\citenamefont {Guimarães}, \citenamefont {Lim}, \citenamefont {Vasilevskiy}, \citenamefont {Huelga},\ and\ \citenamefont {Plenio}}]{guimaraes2023noiseassisted}%
  \BibitemOpen
  \bibfield  {author} {\bibinfo {author} {\bibfnamefont {J.~D.}\ \bibnamefont {Guimarães}}, \bibinfo {author} {\bibfnamefont {J.}~\bibnamefont {Lim}}, \bibinfo {author} {\bibfnamefont {M.~I.}\ \bibnamefont {Vasilevskiy}}, \bibinfo {author} {\bibfnamefont {S.~F.}\ \bibnamefont {Huelga}},\ and\ \bibinfo {author} {\bibfnamefont {M.~B.}\ \bibnamefont {Plenio}},\ }\href@noop {} {\bibinfo {title} {Noise-assisted digital quantum simulation of open systems}} (\bibinfo {year} {2023}),\ \Eprint {https://arxiv.org/abs/2302.14592} {arXiv:2302.14592 [quant-ph]} \BibitemShut {NoStop}%
\bibitem [{\citenamefont {Motta}\ \emph {et~al.}(2019)\citenamefont {Motta}, \citenamefont {Sun}, \citenamefont {Tan}, \citenamefont {O'Rourke}, \citenamefont {Ye}, \citenamefont {Minnich}, \citenamefont {Brand{\~{a}}o},\ and\ \citenamefont {Chan}}]{Motta2019}%
  \BibitemOpen
  \bibfield  {author} {\bibinfo {author} {\bibfnamefont {M.}~\bibnamefont {Motta}}, \bibinfo {author} {\bibfnamefont {C.}~\bibnamefont {Sun}}, \bibinfo {author} {\bibfnamefont {A.~T.~K.}\ \bibnamefont {Tan}}, \bibinfo {author} {\bibfnamefont {M.~J.}\ \bibnamefont {O'Rourke}}, \bibinfo {author} {\bibfnamefont {E.}~\bibnamefont {Ye}}, \bibinfo {author} {\bibfnamefont {A.~J.}\ \bibnamefont {Minnich}}, \bibinfo {author} {\bibfnamefont {F.~G. S.~L.}\ \bibnamefont {Brand{\~{a}}o}},\ and\ \bibinfo {author} {\bibfnamefont {G.~K.-L.}\ \bibnamefont {Chan}},\ }\bibfield  {title} {\bibinfo {title} {Determining eigenstates and thermal states on a quantum computer using quantum imaginary time evolution},\ }\href {https://doi.org/10.1038/s41567-019-0704-4} {\bibfield  {journal} {\bibinfo  {journal} {Nature Physics}\ }\textbf {\bibinfo {volume} {16}},\ \bibinfo {pages} {205} (\bibinfo {year} {2019})}\BibitemShut {NoStop}%
\bibitem [{\citenamefont {Xie}\ \emph {et~al.}(2023)\citenamefont {Xie}, \citenamefont {Xue},\ and\ \citenamefont {Zhang}}]{xie2023variational}%
  \BibitemOpen
  \bibfield  {author} {\bibinfo {author} {\bibfnamefont {X.-D.}\ \bibnamefont {Xie}}, \bibinfo {author} {\bibfnamefont {Z.-Y.}\ \bibnamefont {Xue}},\ and\ \bibinfo {author} {\bibfnamefont {D.-B.}\ \bibnamefont {Zhang}},\ }\href@noop {} {\bibinfo {title} {Variational quantum eigensolvers for the non-hermitian systems by variance minimization}} (\bibinfo {year} {2023}),\ \Eprint {https://arxiv.org/abs/2305.19807} {arXiv:2305.19807 [quant-ph]} \BibitemShut {NoStop}%
\bibitem [{\citenamefont {Zhang}\ \emph {et~al.}(2022)\citenamefont {Zhang}, \citenamefont {Chen}, \citenamefont {Yuan},\ and\ \citenamefont {Yin}}]{Zhang_2022}%
  \BibitemOpen
  \bibfield  {author} {\bibinfo {author} {\bibfnamefont {D.-B.}\ \bibnamefont {Zhang}}, \bibinfo {author} {\bibfnamefont {B.-L.}\ \bibnamefont {Chen}}, \bibinfo {author} {\bibfnamefont {Z.-H.}\ \bibnamefont {Yuan}},\ and\ \bibinfo {author} {\bibfnamefont {T.}~\bibnamefont {Yin}},\ }\bibfield  {title} {\bibinfo {title} {Variational quantum eigensolvers by variance minimization},\ }\href {https://doi.org/10.1088/1674-1056/ac8a8d} {\bibfield  {journal} {\bibinfo  {journal} {Chinese Physics B}\ }\textbf {\bibinfo {volume} {31}},\ \bibinfo {pages} {120301} (\bibinfo {year} {2022})}\BibitemShut {NoStop}%
\bibitem [{\citenamefont {Liu}\ \emph {et~al.}(2023)\citenamefont {Liu}, \citenamefont {Yang}, \citenamefont {Tang}, \citenamefont {Che}, \citenamefont {Nie}, \citenamefont {Xin}, \citenamefont {Li},\ and\ \citenamefont {Lu}}]{Liu_2023}%
  \BibitemOpen
  \bibfield  {author} {\bibinfo {author} {\bibfnamefont {H.}~\bibnamefont {Liu}}, \bibinfo {author} {\bibfnamefont {X.}~\bibnamefont {Yang}}, \bibinfo {author} {\bibfnamefont {K.}~\bibnamefont {Tang}}, \bibinfo {author} {\bibfnamefont {L.}~\bibnamefont {Che}}, \bibinfo {author} {\bibfnamefont {X.}~\bibnamefont {Nie}}, \bibinfo {author} {\bibfnamefont {T.}~\bibnamefont {Xin}}, \bibinfo {author} {\bibfnamefont {J.}~\bibnamefont {Li}},\ and\ \bibinfo {author} {\bibfnamefont {D.}~\bibnamefont {Lu}},\ }\bibfield  {title} {\bibinfo {title} {Practical quantum simulation of small-scale non-hermitian dynamics},\ }\bibfield  {journal} {\bibinfo  {journal} {Physical Review A}\ }\textbf {\bibinfo {volume} {107}},\ \href {https://doi.org/10.1103/physreva.107.062608} {10.1103/physreva.107.062608} (\bibinfo {year} {2023})\BibitemShut {NoStop}%
\bibitem [{\citenamefont {Childs}\ and\ \citenamefont {Wiebe}(2012)}]{LCU}%
  \BibitemOpen
  \bibfield  {author} {\bibinfo {author} {\bibfnamefont {A.~M.}\ \bibnamefont {Childs}}\ and\ \bibinfo {author} {\bibfnamefont {N.}~\bibnamefont {Wiebe}},\ }\bibfield  {title} {\bibinfo {title} {Hamiltonian simulation using linear combinations of unitary operations},\ }\href@noop {} {\ \textbf {\bibinfo {volume} {12}},\ \bibinfo {pages} {901–924} (\bibinfo {year} {2012})}\BibitemShut {NoStop}%
\bibitem [{\citenamefont {Schlimgen}\ \emph {et~al.}(2022)\citenamefont {Schlimgen}, \citenamefont {Head-Marsden}, \citenamefont {Sager-Smith}, \citenamefont {Narang},\ and\ \citenamefont {Mazziotti}}]{schlimgen2022quantum}%
  \BibitemOpen
  \bibfield  {author} {\bibinfo {author} {\bibfnamefont {A.~W.}\ \bibnamefont {Schlimgen}}, \bibinfo {author} {\bibfnamefont {K.}~\bibnamefont {Head-Marsden}}, \bibinfo {author} {\bibfnamefont {L.~M.}\ \bibnamefont {Sager-Smith}}, \bibinfo {author} {\bibfnamefont {P.}~\bibnamefont {Narang}},\ and\ \bibinfo {author} {\bibfnamefont {D.~A.}\ \bibnamefont {Mazziotti}},\ }\href@noop {} {\bibinfo {title} {Quantum simulation of open quantum systems using density-matrix purification}} (\bibinfo {year} {2022}),\ \Eprint {https://arxiv.org/abs/2207.07112} {arXiv:2207.07112 [quant-ph]} \BibitemShut {NoStop}%
\bibitem [{\citenamefont {Wei}\ \emph {et~al.}(2016)\citenamefont {Wei}, \citenamefont {Ruan},\ and\ \citenamefont {Long}}]{Wei2016}%
  \BibitemOpen
  \bibfield  {author} {\bibinfo {author} {\bibfnamefont {S.-J.}\ \bibnamefont {Wei}}, \bibinfo {author} {\bibfnamefont {D.}~\bibnamefont {Ruan}},\ and\ \bibinfo {author} {\bibfnamefont {G.-L.}\ \bibnamefont {Long}},\ }\bibfield  {title} {\bibinfo {title} {Duality quantum algorithm efficiently simulates open quantum systems},\ }\bibfield  {journal} {\bibinfo  {journal} {Scientific Reports}\ }\textbf {\bibinfo {volume} {6}},\ \href {https://doi.org/10.1038/srep30727} {10.1038/srep30727} (\bibinfo {year} {2016})\BibitemShut {NoStop}%
\bibitem [{\citenamefont {Hu}\ \emph {et~al.}(2020)\citenamefont {Hu}, \citenamefont {Xia},\ and\ \citenamefont {Kais}}]{Hu2020}%
  \BibitemOpen
  \bibfield  {author} {\bibinfo {author} {\bibfnamefont {Z.}~\bibnamefont {Hu}}, \bibinfo {author} {\bibfnamefont {R.}~\bibnamefont {Xia}},\ and\ \bibinfo {author} {\bibfnamefont {S.}~\bibnamefont {Kais}},\ }\bibfield  {title} {\bibinfo {title} {A quantum algorithm for evolving open quantum dynamics on quantum computing devices},\ }\bibfield  {journal} {\bibinfo  {journal} {Scientific Reports}\ }\textbf {\bibinfo {volume} {10}},\ \href {https://doi.org/10.1038/s41598-020-60321-x} {10.1038/s41598-020-60321-x} (\bibinfo {year} {2020})\BibitemShut {NoStop}%
\bibitem [{\citenamefont {Hubisz}\ \emph {et~al.}(2021)\citenamefont {Hubisz}, \citenamefont {Sambasivam},\ and\ \citenamefont {Unmuth-Yockey}}]{PhysRevA.104.052420}%
  \BibitemOpen
  \bibfield  {author} {\bibinfo {author} {\bibfnamefont {J.}~\bibnamefont {Hubisz}}, \bibinfo {author} {\bibfnamefont {B.}~\bibnamefont {Sambasivam}},\ and\ \bibinfo {author} {\bibfnamefont {J.}~\bibnamefont {Unmuth-Yockey}},\ }\bibfield  {title} {\bibinfo {title} {Quantum algorithms for open lattice field theory},\ }\href {https://doi.org/10.1103/PhysRevA.104.052420} {\bibfield  {journal} {\bibinfo  {journal} {Phys. Rev. A}\ }\textbf {\bibinfo {volume} {104}},\ \bibinfo {pages} {052420} (\bibinfo {year} {2021})}\BibitemShut {NoStop}%
\bibitem [{\citenamefont {Hu}\ \emph {et~al.}(2022)\citenamefont {Hu}, \citenamefont {Head-Marsden}, \citenamefont {Mazziotti}, \citenamefont {Narang},\ and\ \citenamefont {Kais}}]{Hu2022generalquantum}%
  \BibitemOpen
  \bibfield  {author} {\bibinfo {author} {\bibfnamefont {Z.}~\bibnamefont {Hu}}, \bibinfo {author} {\bibfnamefont {K.}~\bibnamefont {Head-Marsden}}, \bibinfo {author} {\bibfnamefont {D.~A.}\ \bibnamefont {Mazziotti}}, \bibinfo {author} {\bibfnamefont {P.}~\bibnamefont {Narang}},\ and\ \bibinfo {author} {\bibfnamefont {S.}~\bibnamefont {Kais}},\ }\bibfield  {title} {\bibinfo {title} {A general quantum algorithm for open quantum dynamics demonstrated with the {F}enna-{M}atthews-{O}lson complex},\ }\href {https://doi.org/10.22331/q-2022-05-30-726} {\bibfield  {journal} {\bibinfo  {journal} {{Quantum}}\ }\textbf {\bibinfo {volume} {6}},\ \bibinfo {pages} {726} (\bibinfo {year} {2022})}\BibitemShut {NoStop}%
\bibitem [{\citenamefont {Head-Marsden}\ \emph {et~al.}(2021)\citenamefont {Head-Marsden}, \citenamefont {Krastanov}, \citenamefont {Mazziotti},\ and\ \citenamefont {Narang}}]{PhysRevResearch.3.013182}%
  \BibitemOpen
  \bibfield  {author} {\bibinfo {author} {\bibfnamefont {K.}~\bibnamefont {Head-Marsden}}, \bibinfo {author} {\bibfnamefont {S.}~\bibnamefont {Krastanov}}, \bibinfo {author} {\bibfnamefont {D.~A.}\ \bibnamefont {Mazziotti}},\ and\ \bibinfo {author} {\bibfnamefont {P.}~\bibnamefont {Narang}},\ }\bibfield  {title} {\bibinfo {title} {Capturing non-markovian dynamics on near-term quantum computers},\ }\href {https://doi.org/10.1103/PhysRevResearch.3.013182} {\bibfield  {journal} {\bibinfo  {journal} {Phys. Rev. Res.}\ }\textbf {\bibinfo {volume} {3}},\ \bibinfo {pages} {013182} (\bibinfo {year} {2021})}\BibitemShut {NoStop}%
\bibitem [{\citenamefont {Naghiloo}\ \emph {et~al.}(2019)\citenamefont {Naghiloo}, \citenamefont {Abbasi}, \citenamefont {Joglekar},\ and\ \citenamefont {Murch}}]{Naghiloo2019-wh}%
  \BibitemOpen
  \bibfield  {author} {\bibinfo {author} {\bibfnamefont {M.}~\bibnamefont {Naghiloo}}, \bibinfo {author} {\bibfnamefont {M.}~\bibnamefont {Abbasi}}, \bibinfo {author} {\bibfnamefont {Y.~N.}\ \bibnamefont {Joglekar}},\ and\ \bibinfo {author} {\bibfnamefont {K.~W.}\ \bibnamefont {Murch}},\ }\bibfield  {title} {\bibinfo {title} {Quantum state tomography across the exceptional point in a single dissipative qubit},\ }\href@noop {} {\bibfield  {journal} {\bibinfo  {journal} {Nat. Phys.}\ }\textbf {\bibinfo {volume} {15}},\ \bibinfo {pages} {1232} (\bibinfo {year} {2019})}\BibitemShut {NoStop}%
\bibitem [{\citenamefont {Dogra}\ \emph {et~al.}(2021)\citenamefont {Dogra}, \citenamefont {Melnikov},\ and\ \citenamefont {Paraoanu}}]{Dogra2021-ln}%
  \BibitemOpen
  \bibfield  {author} {\bibinfo {author} {\bibfnamefont {S.}~\bibnamefont {Dogra}}, \bibinfo {author} {\bibfnamefont {A.~A.}\ \bibnamefont {Melnikov}},\ and\ \bibinfo {author} {\bibfnamefont {G.~S.}\ \bibnamefont {Paraoanu}},\ }\bibfield  {title} {\bibinfo {title} {Quantum simulation of parity--time symmetry breaking with a superconducting quantum processor},\ }\href@noop {} {\bibfield  {journal} {\bibinfo  {journal} {Commun. Phys.}\ }\textbf {\bibinfo {volume} {4}} (\bibinfo {year} {2021})}\BibitemShut {NoStop}%
\bibitem [{\citenamefont {Lin}\ \emph {et~al.}(2022)\citenamefont {Lin}, \citenamefont {Zhang}, \citenamefont {Long}, \citenamefont {ang Fan}, \citenamefont {Li}, \citenamefont {Tang}, \citenamefont {Li}, \citenamefont {Nie}, \citenamefont {Xin}, \citenamefont {Liu},\ and\ \citenamefont {Lu}}]{Lin2022}%
  \BibitemOpen
  \bibfield  {author} {\bibinfo {author} {\bibfnamefont {Z.}~\bibnamefont {Lin}}, \bibinfo {author} {\bibfnamefont {L.}~\bibnamefont {Zhang}}, \bibinfo {author} {\bibfnamefont {X.}~\bibnamefont {Long}}, \bibinfo {author} {\bibfnamefont {Y.}~\bibnamefont {ang Fan}}, \bibinfo {author} {\bibfnamefont {Y.}~\bibnamefont {Li}}, \bibinfo {author} {\bibfnamefont {K.}~\bibnamefont {Tang}}, \bibinfo {author} {\bibfnamefont {J.}~\bibnamefont {Li}}, \bibinfo {author} {\bibfnamefont {X.}~\bibnamefont {Nie}}, \bibinfo {author} {\bibfnamefont {T.}~\bibnamefont {Xin}}, \bibinfo {author} {\bibfnamefont {X.-J.}\ \bibnamefont {Liu}},\ and\ \bibinfo {author} {\bibfnamefont {D.}~\bibnamefont {Lu}},\ }\bibfield  {title} {\bibinfo {title} {Experimental quantum simulation of non-hermitian dynamical topological states using stochastic schr\"{o}dinger equation},\ }\bibfield  {journal} {\bibinfo  {journal} {npj Quantum Information}\ }\textbf {\bibinfo {volume} {8}},\ \href {https://doi.org/10.1038/s41534-022-00587-3}
  {10.1038/s41534-022-00587-3} (\bibinfo {year} {2022})\BibitemShut {NoStop}%
\bibitem [{\citenamefont {Powers}\ \emph {et~al.}(2023)\citenamefont {Powers}, \citenamefont {Bassman~Oftelie}, \citenamefont {Camps},\ and\ \citenamefont {de~Jong}}]{Powers2023-en}%
  \BibitemOpen
  \bibfield  {author} {\bibinfo {author} {\bibfnamefont {C.}~\bibnamefont {Powers}}, \bibinfo {author} {\bibfnamefont {L.}~\bibnamefont {Bassman~Oftelie}}, \bibinfo {author} {\bibfnamefont {D.}~\bibnamefont {Camps}},\ and\ \bibinfo {author} {\bibfnamefont {W.~A.}\ \bibnamefont {de~Jong}},\ }\bibfield  {title} {\bibinfo {title} {Exploring finite temperature properties of materials with quantum computers},\ }\href@noop {} {\bibfield  {journal} {\bibinfo  {journal} {Sci. Rep.}\ }\textbf {\bibinfo {volume} {13}},\ \bibinfo {pages} {1986} (\bibinfo {year} {2023})}\BibitemShut {NoStop}%
\bibitem [{\citenamefont {Shende}\ \emph {et~al.}(2006)\citenamefont {Shende}, \citenamefont {Bullock},\ and\ \citenamefont {Markov}}]{Shende_2006}%
  \BibitemOpen
  \bibfield  {author} {\bibinfo {author} {\bibfnamefont {V.}~\bibnamefont {Shende}}, \bibinfo {author} {\bibfnamefont {S.}~\bibnamefont {Bullock}},\ and\ \bibinfo {author} {\bibfnamefont {I.}~\bibnamefont {Markov}},\ }\bibfield  {title} {\bibinfo {title} {Synthesis of quantum-logic circuits},\ }\href {https://doi.org/10.1109/tcad.2005.855930} {\bibfield  {journal} {\bibinfo  {journal} {{IEEE} Transactions on Computer-Aided Design of Integrated Circuits and Systems}\ }\textbf {\bibinfo {volume} {25}},\ \bibinfo {pages} {1000} (\bibinfo {year} {2006})}\BibitemShut {NoStop}%
\bibitem [{\citenamefont {Viola}\ and\ \citenamefont {Lloyd}(1998)}]{PhysRevA.58.2733}%
  \BibitemOpen
  \bibfield  {author} {\bibinfo {author} {\bibfnamefont {L.}~\bibnamefont {Viola}}\ and\ \bibinfo {author} {\bibfnamefont {S.}~\bibnamefont {Lloyd}},\ }\bibfield  {title} {\bibinfo {title} {Dynamical suppression of decoherence in two-state quantum systems},\ }\href {https://doi.org/10.1103/PhysRevA.58.2733} {\bibfield  {journal} {\bibinfo  {journal} {Phys. Rev. A}\ }\textbf {\bibinfo {volume} {58}},\ \bibinfo {pages} {2733} (\bibinfo {year} {1998})}\BibitemShut {NoStop}%
\bibitem [{\citenamefont {Zanardi}(1999)}]{ZANARDI199977}%
  \BibitemOpen
  \bibfield  {author} {\bibinfo {author} {\bibfnamefont {P.}~\bibnamefont {Zanardi}},\ }\bibfield  {title} {\bibinfo {title} {Symmetrizing evolutions},\ }\href {https://doi.org/https://doi.org/10.1016/S0375-9601(99)00365-5} {\bibfield  {journal} {\bibinfo  {journal} {Physics Letters A}\ }\textbf {\bibinfo {volume} {258}},\ \bibinfo {pages} {77} (\bibinfo {year} {1999})}\BibitemShut {NoStop}%
\bibitem [{\citenamefont {Vitali}\ and\ \citenamefont {Tombesi}(1999)}]{PhysRevA.59.4178}%
  \BibitemOpen
  \bibfield  {author} {\bibinfo {author} {\bibfnamefont {D.}~\bibnamefont {Vitali}}\ and\ \bibinfo {author} {\bibfnamefont {P.}~\bibnamefont {Tombesi}},\ }\bibfield  {title} {\bibinfo {title} {Using parity kicks for decoherence control},\ }\href {https://doi.org/10.1103/PhysRevA.59.4178} {\bibfield  {journal} {\bibinfo  {journal} {Phys. Rev. A}\ }\textbf {\bibinfo {volume} {59}},\ \bibinfo {pages} {4178} (\bibinfo {year} {1999})}\BibitemShut {NoStop}%
\bibitem [{\citenamefont {Duan}\ and\ \citenamefont {Guo}(1999)}]{DUAN1999139}%
  \BibitemOpen
  \bibfield  {author} {\bibinfo {author} {\bibfnamefont {L.-M.}\ \bibnamefont {Duan}}\ and\ \bibinfo {author} {\bibfnamefont {G.-C.}\ \bibnamefont {Guo}},\ }\bibfield  {title} {\bibinfo {title} {Suppressing environmental noise in quantum computation through pulse control},\ }\href {https://doi.org/https://doi.org/10.1016/S0375-9601(99)00592-7} {\bibfield  {journal} {\bibinfo  {journal} {Physics Letters A}\ }\textbf {\bibinfo {volume} {261}},\ \bibinfo {pages} {139} (\bibinfo {year} {1999})}\BibitemShut {NoStop}%
\bibitem [{\citenamefont {Viola}\ \emph {et~al.}(1999)\citenamefont {Viola}, \citenamefont {Knill},\ and\ \citenamefont {Lloyd}}]{PhysRevLett.82.2417}%
  \BibitemOpen
  \bibfield  {author} {\bibinfo {author} {\bibfnamefont {L.}~\bibnamefont {Viola}}, \bibinfo {author} {\bibfnamefont {E.}~\bibnamefont {Knill}},\ and\ \bibinfo {author} {\bibfnamefont {S.}~\bibnamefont {Lloyd}},\ }\bibfield  {title} {\bibinfo {title} {Dynamical decoupling of open quantum systems},\ }\href {https://doi.org/10.1103/PhysRevLett.82.2417} {\bibfield  {journal} {\bibinfo  {journal} {Phys. Rev. Lett.}\ }\textbf {\bibinfo {volume} {82}},\ \bibinfo {pages} {2417} (\bibinfo {year} {1999})}\BibitemShut {NoStop}%
\bibitem [{\citenamefont {Charles}\ \emph {et~al.}(2023)\citenamefont {Charles}, \citenamefont {Gustafson}, \citenamefont {Hardt}, \citenamefont {Herren}, \citenamefont {Hogan}, \citenamefont {Lamm}, \citenamefont {Starecheski}, \citenamefont {de~Water},\ and\ \citenamefont {Wagman}}]{charles2023simulating}%
  \BibitemOpen
  \bibfield  {author} {\bibinfo {author} {\bibfnamefont {C.}~\bibnamefont {Charles}}, \bibinfo {author} {\bibfnamefont {E.~J.}\ \bibnamefont {Gustafson}}, \bibinfo {author} {\bibfnamefont {E.}~\bibnamefont {Hardt}}, \bibinfo {author} {\bibfnamefont {F.}~\bibnamefont {Herren}}, \bibinfo {author} {\bibfnamefont {N.}~\bibnamefont {Hogan}}, \bibinfo {author} {\bibfnamefont {H.}~\bibnamefont {Lamm}}, \bibinfo {author} {\bibfnamefont {S.}~\bibnamefont {Starecheski}}, \bibinfo {author} {\bibfnamefont {R.~S.~V.}\ \bibnamefont {de~Water}},\ and\ \bibinfo {author} {\bibfnamefont {M.~L.}\ \bibnamefont {Wagman}},\ }\href@noop {} {\bibinfo {title} {Simulating $\mathbb{Z}_2$ lattice gauge theory on a quantum computer}} (\bibinfo {year} {2023}),\ \Eprint {https://arxiv.org/abs/2305.02361} {arXiv:2305.02361 [hep-lat]} \BibitemShut {NoStop}%
\bibitem [{\citenamefont {Gustafson}\ \emph {et~al.}()\citenamefont {Gustafson}, \citenamefont {Hite}, \citenamefont {Hubisz}, \citenamefont {Sambasivam},\ and\ \citenamefont {Unmuth-Yokey}}]{GithubRepo}%
  \BibitemOpen
  \bibfield  {author} {\bibinfo {author} {\bibfnamefont {E.}~\bibnamefont {Gustafson}}, \bibinfo {author} {\bibfnamefont {M.}~\bibnamefont {Hite}}, \bibinfo {author} {\bibfnamefont {J.}~\bibnamefont {Hubisz}}, \bibinfo {author} {\bibfnamefont {B.}~\bibnamefont {Sambasivam}},\ and\ \bibinfo {author} {\bibfnamefont {J.}~\bibnamefont {Unmuth-Yokey}},\ }\href {https://github.com/TheBharatheon/NHIsingAlt.git} {\bibinfo {title} {{Github-repository for Quantum simulation of non-Hermitian 1D Ising model}}}\BibitemShut {NoStop}%
\bibitem [{\citenamefont {A~Rahman}\ \emph {et~al.}(2022)\citenamefont {A~Rahman}, \citenamefont {Lewis}, \citenamefont {Mendicelli},\ and\ \citenamefont {Powell}}]{PhysRevD.106.074502}%
  \BibitemOpen
  \bibfield  {author} {\bibinfo {author} {\bibfnamefont {S.}~\bibnamefont {A~Rahman}}, \bibinfo {author} {\bibfnamefont {R.}~\bibnamefont {Lewis}}, \bibinfo {author} {\bibfnamefont {E.}~\bibnamefont {Mendicelli}},\ and\ \bibinfo {author} {\bibfnamefont {S.}~\bibnamefont {Powell}},\ }\bibfield  {title} {\bibinfo {title} {Self-mitigating trotter circuits for su(2) lattice gauge theory on a quantum computer},\ }\href {https://doi.org/10.1103/PhysRevD.106.074502} {\bibfield  {journal} {\bibinfo  {journal} {Phys. Rev. D}\ }\textbf {\bibinfo {volume} {106}},\ \bibinfo {pages} {074502} (\bibinfo {year} {2022})}\BibitemShut {NoStop}%
\end{thebibliography}%

\end{document}